 \documentclass[manuscript]{emulateapj}

\usepackage{natbib}
\usepackage{lscape}
\usepackage{rotating}
\usepackage{bigstrut}

\bibpunct{(}{)}{,}{a}{}{,}

\slugcomment{ApJ \textit{in press}}

\shorttitle{Dark matter in galaxy groups}
\shortauthors{Thanjavur, Crampton  \& Willis}

\begin{document}
\def\asec{^{\prime\prime}}
\def\mag{\:{\mathrm mag}}
\def\angs{\mathrm{\AA}}
\def\sqdeg{\:\mathrm{deg^2}}
\def\deg{\ensuremath{^\circ}}
\def\omegam{$\Omega_{\mathrm m}$}
\def\omegab{$\Omega_{\mathrm b}$}
\def\omegabh{$\Omega_{\mathrm b} h^2$}
\def\omegal{$\Omega_{\Lambda}$}
\def\ovrdn{$\Delta_{200}$}
\def\ovrdnr{$\Delta(r)$}
\def\ovrdnv{$\Delta_{vir}$}
\def\vcr{$v_{\mathrm c}(r)$}
\def\vc{$v_{\mathrm c}(r)$}
\def\halpha{$\mathrm H_{\alpha}$}
\def\rhoc{$\rho_{\mathrm crit}$}
\def\rhocore{$\rho_{0}$}
\def\rcore{$r_{\mathrm c}$}
\def\rhobar{$\bar{\rho}$}
\def\rhor{$\rho(r)$}
\def\mr{$M(r)$}
\def\radius{$r$}
\def\rscale{$r_{\mathrm s}$}
\def\deltac{$\delta_{\mathrm c}$}
\def\lcdm{$\Lambda$CDM }
\def\scdm{$S$CDM }
\def\rnfw{$r^{-1}$}
\def\mathrmoore{$r^{-1.5}$}
\def\hub{$h$}
\newcommand{\mc}[3]{\multicolumn{#1}{#2}{#3}}
\def\G{{\mathrm G}}
\def\c{{\mathrm c}}
\def\gsim{ \lower .75ex \hbox{$\sim$} \llap{\raise .27ex \hbox{$>$}} }
\def\lsim{ \lower .75ex \hbox{$\sim$} \llap{\raise .27ex \hbox{$<$}} }
\def\Mpc{{\mathrm Mpc}}
\def\kpc{{\mathrm kpc}}
\def\pc{{\mathrm pc}}
\def\Lsun{{\mathrm L_\odot}}
\def\Msun{\:{\rm M_\odot}}
\def\gsim{\ga}
\def\eg{e.g.\ }
\def\lsim{\la}
\def\etal{{et al.\ }}
\def\mpc {\mathrm{Mpc}}
\def\kpc {\mathrm{kpc}}
\def\msun {\:{\rm M_\odot}}
\def\ergs {{\mathrm erg} \, {\mathrm s}^{-1}}
\def\cm {{\mathrm cm}}
\def\kms {{\mathrm {\,km \, s^{-1}}}}
\def\Hz {{\mathrm Hz}}
\def\yr {{\mathrm yr}}
\def\gyr {{\mathrm Gyr}}
\def\arcmin {{\mathrm arcmin}}
\def\G {{ G  }}
\def\magasec {\mathrm {\, mag.arc.sec^{-2}}} 
\def\lsim{\mathrel{\hbox{\rlap{\hbox{\lower4pt\hbox{$\sim$}}}\hbox{$<$}}}}
\def\gsim{\mathrel{\hbox{\rlap{\hbox{\lower4pt\hbox{$\sim$}}}\hbox{$>$}}}}

\def\and   {\mathrm {et al.} \mathrm}  
\def\mathrmd {\mathrm d}



\title{DARK MATTER DISTRIBUTION IN GALAXY GROUPS FROM COMBINED STRONG LENSING AND DYNAMICS ANALYSIS}

\author{Karun Thanjavur\altaffilmark{1,2,3}, David Crampton\altaffilmark{2,3},
	and Jon Willis\altaffilmark{2}}

\altaffiltext{2}{Canada France Hawaii Telescope, 65-1238 Mamalahoa Hwy, 
        Kamuela, Hawaii 96743, USA; karun@cfht.hawaii.edu}
\altaffiltext{2}{Department of Physics \& Astronomy, University of Victoria, 
        Victoria, BC, V8P 1A1, Canada}
\altaffiltext{3}{National Research Council of Canada, Herzberg Institute of 
        Astrophysics, 5071 West Saanich Road, Victoria, BC, V9E 2E7, 
        Canada}

\begin{abstract}

Using a combined analysis of strong lensing and galaxy dynamics, we characterize the mass distributions and the mass-to-light (M/L) ratios of galaxy groups, virialized structures in the mass range of few $\times\,10^{14}\,\msun$, which form an important transition regime in the hierarchical assembly of mass in $\Lambda$CDM cosmology. Our goals are to not only map the mass distributions, but to also test whether the underlying density distribution at this mass scale is dark matter dominated, NFW-like as hypothesized by the standard cosmogony, or isothermal as observed in baryon rich massive field galaxies. We present details of our lensing + galaxy dynamics formalism built around three representative density profiles, the dark matter dominant NFW and Hernquist distributions, compared with the Softened Isothermal Sphere which matches baryon rich galaxy scale objects. By testing the effects on the characteristics of these distributions due to variations in their parameters, we show that mass measurements in the core of the group ($r/r_{vir}\sim0.2$), determined jointly from a lens model and from differential velocity dispersion estimates, may effectively distinguish between these density distributions. We apply our method to MOS observations of two groups, SL2SJ143000+554648 and SL2SJ143139+553323, drawn from our catalog of galaxy group-scale lenses discovered in CFHTLS-Wide imaging. With the lensing and dynamical mass estimates from our observations along with a maximum likelihood estimator built around our model, we estimate the concentration index characterizing each density distribution and the corresponding virial mass of each group. Our likelihood estimation indicates that both groups are dark matter dominant, and rejects the isothermal distribution at $\gg3\sigma$ level. For both groups, the estimated $i$-band M/L ratios of $\sim260\,\msun \Lsun^{-1}$, is similar to other published values for groups. The Gaussian distribution of the velocities of their member galaxies support a high degree of virialization. The differences in their virial masses, 2.8 and 1.6$\times10^{14}\msun$, and velocity dispersions, 720 and 560$\kms$, may indicate however that each group is at a different stage of transition to a cluster. We aim to populate this important transition regime with additional results from ongoing observations of the remaining lensing groups in our catalog    

\end{abstract}



\section{Introduction} \label{review}

\indent In the hierarchical assembly of mass in the standard $\Lambda$CDM cosmogony, galaxy groups with typical masses, M $\sim$few$\times10^{14}\msun$, form an important transition stage in the build up of galaxy clusters (with M$\sim 10^{15}\msun$) from the basic building blocks, the field galaxy population with $M\sim\,10^{12}\msun$. In clusters, which lie at the upper end of this hierarchy of virialized objects, structure evolution is driven primarily by the gravitational physics governing dark matter interactions - in clusters, with observed mass-to-light (M/L) ratios of a several hundred, the baryon fraction is too small to play any significant role. Consequently, their observed mass distributions are well fit by NFW \citep{Nava97} or similar profiles which describe dark matter distributions \citep{Lope09, Broa08, Tu08, Come07, Rine06, Smit05a}. On the scale of galaxies, however, the observed M/L $\leq$10 indicates that baryons contribute a significant fraction of their mass. The rapid cooling of baryons and condensation to the galaxy cores leads to more isothermal density profiles \citep{Gava08, Koop06, Treu04, Free01}. Whether galaxy groups, which lie intermediate in the mass scale between these two classes of structures, are dark matter dominated NFW-like systems, or if they retain the isothermal signatures of their dominant central galaxies is an important but unresolved question on which we focus our investigations. 

\indent In the Cluster Infall Regions in the Sloan Digital Sky (CIRS) survey of 72 low redshift clusters, \citet{Rine06} found that the mass distributions are well fit by \emph{spherically symmetric} NFW \citep{Nava97} or Hernquist profiles \citep{Hern90}, popular parametric representations of dark matter distribution. Recent results of mass estimates using galaxy dynamics from detailed spectroscopy of 48 low redshift ($z < 0.1$) clusters, along with mass estimates from their X-ray luminosities also show that cluster masses are dark matter dominated \citep{Cava09}. \citet{Lope09} extend these cluster mass estimates up to $z\sim0.5$ with their NoSOCS-SDSS sample of 7400 systems. With strong lensing analysis, \citet{Broa08, Tu08, Come07} and \citet{Smit05a} independently show that the slope of the cluster density profiles match the $\Lambda$CDM prediction. From these complementary observational results, it is clear that the structure of clusters is governed by the dominant dark matter content.       

\indent For galaxies, the observed M/L$\leq$10, indicates that the visible baryonic matter, gas and stars, contributes a significant fraction of the mass; early type galaxies show a higher M/L ($\sim10$) value \citep{Free01}, indicating a larger dark matter component than their spiral galaxy counterparts. Even so, results from a joint strong lensing and stellar dynamics analysis for 15 early type galaxies at z $\leq 0.33$ from the Strong Lensing in Advanced Camera for Surveys, SLACS, \citep{Koop06}, show that their density drops off inversely with the square of the radius, $\rho(r)\,\propto\, r^{-2}$, indicative of an \emph{isothermal} instead of NFW-like distribution. Within the Einstein radius, the observed contribution of dark matter to the total mass in these SLACS sample is only $\sim25\%$, indicating that the inner core is baryon rich while the dark matter is dominant only at larger radii. These results are consistent with results for 5 early type field galaxies at higher redshift ($0.5 \leq z \leq 1$) in the Lenses Structure and Dynamics survey \citep{Treu04}. In addition, the lens model for the spectacular double Einstein ring, SDSS J0946+1006, where the deflector is an early type galaxy at z=0.22, gives similar values for the power law slope of the density profile as well as ellipticity \citep{Gava08}, indicating that the distributions of mass at the scale of galaxies are clearly different from the NFW-like distributions observed in clusters. The interaction of baryons leading to rapid cooling and condensation are the likely cause of these observed differences.

\indent In the $\Lambda$CDM picture of hierarchical mass assembly, galaxy groups are built from field galaxies, such as those studied in SLACS. The density profile of the resulting group may then be expected to match that of the dominant central galaxy or appear as a smoothed out version of the contributions of all the group members. An isothermal mass density profile is indeed shown in a recent detailed analysis of CLASS B2108+213, a galaxy group at z=0.365, using lensing and galaxy kinematics, though the non-Gaussian velocity distribution indicates that the group is not virialized \citep{McKe09}. With strong lensing analysis of 5 groups discovered in SDSS imaging, \citet{Kubo09} obtain singular isothermal velocity dispersions in the range, $\sigma_{SIS}\sim464-882\kms$. Similarly, with joint strong and weak lensing analysis, \citet{Limo09} derive $\sigma_{SIS}\sim500\kms$ for 13 galaxy groups taken from the SL2S\footnote{Strong Lensing Legacy Survey, \citep{Caba07}} catalog. With stacked weak lensing, \citet{Hoek01} obtain $\sigma_{SIS}\sim274\kms$ for the best fit isothermal sphere to the less massive groups in the CNOC2\footnote{Canadian Network for Observational Cosmology Field Galaxy Redshift Survey} survey.   

\indent On the other hand, $\Lambda$CDM also predicts that the dark matter distribution should be self similar over a wide range of virialised masses \citep{Nava97}, with groups representing the lower mass end of clusters. \citet{Park05} do indeed find that the weak lensing mass profiles of 116 CNOC2 galaxy groups are fit equally well by NFW and isothermal spheres. If groups are scaled down versions of clusters with NFW-like mass profiles, observed scaling relations for galaxy groups, such as the X-ray luminosity-temperature ($L_X$ - T) relation, should resemble those of clusters. For local clusters, the observed $L\propto T^\alpha$ relation yields $\alpha=2.64 \pm 0.27$ \citep{Mark98}. Interestingly, this value is much higher than the CDM prediction, $\alpha=2$, which does not take into account radiative cooling of the intracluster gas or heating by supernovae and AGN feedback. For groups, \citet{Hels00} find a \emph{much steeper} slope of $\alpha=4.9\ \pm 0.8$ for a sample 24 X-ray bright galaxy groups, though \citet{Mulc04} cautions that aperture effects may play a role in this observed difference. Results from the Group Evolution Multiwavelength Study (GEMS) \citep{Osmo04} and a compilation from several galaxy group surveys \citep{Fass08} find a smaller difference in slope. In addition, large hydrodynamical simulations indicate that merger induced star formation has a strong effect on X-ray luminosities. The active merger of galaxies in group environments may therefore lead to the observed steeper slope than that observed in clusters, as in the case of the lensing group CLASS B2108+213 \citep{McKe09}. 

\indent  All these results underline the pressing need to understand how the mass distribution and M/L ratio of groups fit within the evolution from galaxy to cluster mass scales. These structures, where $\sim55\%$ of all galaxies reside \citep{Eke04a} have masses representative of the critical break between the dark matter dominated clusters and the baryon rich field galaxy population, seen in both observations \citep{Fass08, Mulc04, Mari02, Hels00}, and in detailed hydrodynamical simulations \citep{Hart08}. The M/L ratio of groups is also strongly influenced by ongoing episodes of star formation triggered by the active mergers in the group environment. In a model comparison of the properties of virialised objects of different halo masses in various cosmologies, \citet{Mari02} find that, for $\Lambda$CDM, the M/L ratio increases monotonically with X-ray luminosity as $L_X^{0.5}$ for the mass range running from poor groups, M=$10^{13}\msun$, to that of rich clusters, $10^{15}\msun$. Interestingly, their models also indicate a clear break in the power law relation between optical and X-ray luminosities at the mass scale of a poor group, (consisting of $\leq 5\;M_*$, Milky Way size galaxies). Whether this break is actually seen in observations will be interesting to determine and interpret. 

\indent These unresolved issues related to the mass distribution in galaxy groups have initiated several large observational surveys, e.g,  GEMS \citep{Osmo04}, X-ray Multi-Mirror Large Scale Structure Survey, XMM-LSS \citep{Will05}, XMM/IMACS (XI) Group Project \citep{Rasm06} as well as studies using galaxy group catalogs detected in public wide field surveys, such as the SDSS DR5 \citep{Tago08}, the 2dF \citep{Tago06} and the SL2S catalogs of groups with strong lensing features (referred hence as `lensing groups') detected in CFHTLS Wide imaging \citep{Caba07} and \citep{Limo09}. Using the SDSS DR4-MaxBCG cluster catalog \citep{Koes07}, \citet{Shel09b} have applied the cross correlation weak lensing method to estimate the excess mass-to-light ratio, $\Delta\!M/\Delta\!L$, in these virialized systems, measured relative to the background (field) value. The impressive sample size of $\sim130,000$ structures available in this catalog spans the mass scale from poor groups of a few $10^{12}\msun$, to rich, $10^{15}\msun$ clusters in the redshift interval 0.1 to 0.3 \citep{Koes07}. The stacked sample in each richness bin provides adequate S/N to trace the trend in the M/L ratio with radius extending from a few $\kpc$ in the core to a few tens of $\mpc$, spanning several times the size of a cluster virial radius.  Of particular relevance to our work is their result that the excess M/L correlates with the virial mass as a power law with a slope of $0.33\pm0.02$. By integrating the radial trend up to the virial radius, $r_{200}$, \citet{Shel09b} also provide detailed tables of the M/L ratios of these structures binned by richness and by optical luminosity. These weak lensing results permit us to compare and benchmark the M/L values obtained by our combined strong lensing and galaxy dynamics approach, as discussed in Section \ref{CompRes}.

\indent  Using the catalog of lensing groups\footnote{All the objects in our sample are referred as \emph{groups} though their actual virial masses are yet to be determined} from \citet{Than09}, our objective is to characterize the density distributions and the M/L ratios of structures at this mass scale. Toward this goal, we have developed a joint analysis that combines mass constraints estimated independently from a lens model and from measured galaxy dynamics. Since the lensed image separations are $\ll 10\asec$ in these moderate redshift groups (median z$\sim0.5$), the lens model effectively constrains the mass distribution within the core (radius $\ll 100\kpc$)  of the group. This is complemented well by the dynamical mass estimate from the velocity dispersion of the member galaxies, which is sensitive to the mass at the scale of the virial radius ($\sim1\mpc$). The power of this joint approach has been well illustrated by other applications such as the characterization of early type galaxies in SLACS \citep{Koop06, Treu04}. and of the lensing group, CLASS B2108+213 \citep{McKe09}, 

\indent In Section \ref{LOScomp}, we set forth details of our formalism which relates an assumed underlying density distribution to the projected mass and the line of sight velocity dispersion (LOSVD) measured within an aperture - these are our \emph{principal observables}. The goal is to verify how well observations can distinguish one density distribution from another, using only galaxy dynamics (in \S \ref{LOSres1}), refined in \S \ref{LOSres2} with the inclusion of the lensing mass. We test our approach on the two lensing groups from our catalog for which observations are complete. Section \ref{DMlensObs} describes the instrument configuration, data reduction details, and the observed LOSVD of the member galaxies; the lens model and resulting mass estimates are given in \S \ref{LOSres3}, while the maximum likelihood estimator used to extract the parameters of our model from our observations is described in \S \ref{Modfit}. Section \ref{CompRes} discusses our results in comparison with those from other similar investigations. Section \ref{DC} summarizes our work along with plans for future refinements. \\


\section {Model estimates of projected mass and LOSVD for specific density profiles} \label{LOScomp}

\indent For this analysis, we have chosen two parametric profiles commonly used to describe dark matter distributions, namely the Navarro, Frenk and White, \emph{NFW} \citep{Nava97} and the Hernquist, \emph{HRQ} \citep{Hern90} profiles, which we benchmark against the \emph{Softened} Isothermal Sphere\footnote{The Softened Isothermal Sphere has a non-zero core radius to prevent the singular behavior at r=0 of the \emph{Singular} Isothermal Sphere. In our description, the abbreviation, SIS, consistently represents a \emph{Softened} Isothermal Sphere only}, \emph{SIS}, a generic model used to describe virialized mass distribution \citep{Binn87}. Our choice of the NFW and HRQ profiles for this comparison is motivated by observational results such as those from the CIRS \citep{Rine06} cited in Section \ref{review}. We benchmark these against the SIS, which describes galaxy scale mass distributions. We do not test alternate density parametrizations such as the Kravtsov model \citep{Krav98}, which differ from the NFW profile only in the inner slope at radii less than a few percent of the virial radius; this difference has little effect on the LOSVD measured up to the virial radius. Finally, in this comparison, we assume spherical symmetry for all three profiles and neglect any effects due to triaxiality; this analysis may have to be refined if our observations indicate non-isotropic velocity distributions.  

\indent The standard representation of the three density profiles involve different sets of parameters, namely   the scale radius, $r_s$, for NFW \citep{Nava97}, the scale radius, $a$, for HRQ \citep{Hern90}, and the core radius, $r_c$, of SIS \citep{Binn87}; in addition, a second parameter is used for the normalization. In order to ensure a uniform  comparison of their properties, we first recast them into a non-dimensional form with a consistent set of two parameters. The parameters chosen are the concentration index, $c$, of the profile and the mass enclosed within a prescribed radius. In \S \ref{LOSres1}, where we compare the characteristics of the three profiles against each other, we use the virial mass of a typical rich cluster, $M_{v,200} = 10^{15}\Msun$, for the normalization (with an assumed overdensity parameter, $\Delta=200$ expressed in units of the present critical density of the unverse, $\rho_0^c$). This cluster scale mass was chosen only so that our formalism may be verified by comparing the NFW results against published values in \citet{Loka01} and \citet{Will00a} for a halo of the same virial mass; group scale masses, more pertinent to our work, are used in the second comparison, discussed in \S \ref{LOSres2}. 

\begin{figure*}[htbp] 
  \centering
\includegraphics[width=5.85in, angle=0]{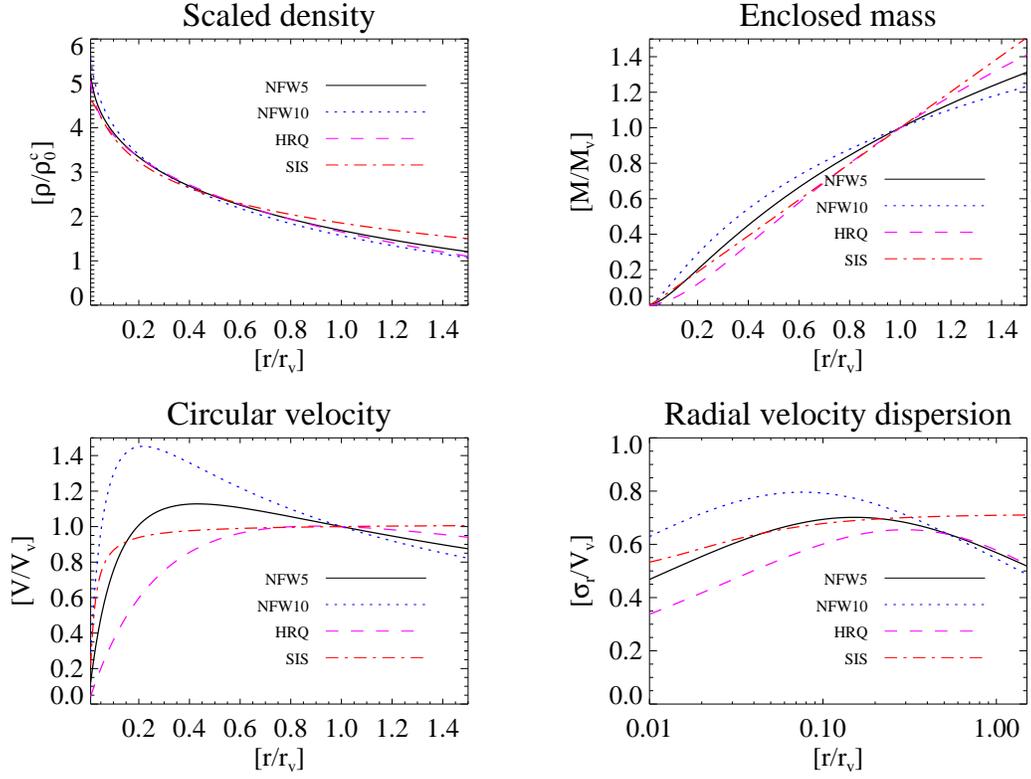} 
  \caption[Comparison of halo properties for NFW, HRQ and SIS density profiles]{Comparison of properties of dark matter halos described by NFW, HRQ and SIS density profiles; for this plot, the profiles have been matched as described in Section \ref{LOScomp}. The top panels show the scaled density, $log(\rho / \rho_0^c)$ (left) and the enclosed mass, $M/M_v$ (right) as functions of the scaled radius, $r/r_v$, while the corresponding bottom panels are the circular velocity, $V/V_v$ and the radial velocity dispersion, $\sigma_r/V_v$.}
   \label{Fig1}
\end{figure*}

\indent In terms of these two parameters, the \emph{non-dimensional} density profiles may be expressed as    
\begin{equation}
\label{eqn5}
\eta(s) = \frac{\rho(s)}{\rho_0^c \: \delta_c} 
 \end{equation}
where the scaled radius, $s=r/r_v$, with $r_v$ being the virial radius. The function, $\eta(s)$, which describes the shape of the profile, is represented for the NFW, HRQ and SIS by,
\begin{equation}
\label{eqn6}
\eta(s) =  \frac{1}{(cs)(1+cs)^2} \qquad \mbox{NFW}
 \end{equation}

\begin{equation}
\label{eqn7}
\eta(s) =  \frac{1}{(cs)(1+cs)^3} \qquad \mbox{HRQ}
 \end{equation}

\begin{equation}
\label{eqn8}
\eta(s) =  \frac{1}{1+(cs)^2} \qquad \mbox{SIS}
 \end{equation}
where, the \emph{concentration index}, $c$, represents $(r_v/r_s)$ for NFW, $(r_v/a)$ for HRQ and $(r_v/r_c)$ for the SIS profiles. The functional form of the \emph{concentration parameter}, $\delta_c $, which contains the normalization scale for each profile, is obtained by setting the expression for the enclosed mass within the virial radius, $M(r_v)$, equal to the virial mass, $M_v$
\begin{equation}
\label{eqn9}
M(r) = \left( \int_0^r 4\pi k^2 \rho(k) dk \right) \Big|_{r=r_v} =  \frac{4}{3}\pi r_v^3(\Delta\:\rho_0^c) \end{equation}
Substituting the normalized density profiles, Equations \ref{eqn5} - \ref{eqn8} and carrying out the integral in Equation \ref{eqn9}, yields analytical expressions for $\delta_c $ for the NFW, HRQ and SIS as,
\begin{equation}
\label{eqn10}
\delta_c =  \frac{1}{3} \frac{\Delta c^3}{\left( \mathrm{ln}(1+c)\:-\: \frac{c}{(1+c)} \right)} \qquad \mbox{NFW}
 \end{equation}

\begin{equation}
\label{eqn11}
\delta_c =  \frac{2}{3} (\Delta c^3) \left( 1+ \frac{1}{c} \right)^2 \qquad \mbox{HRQ}
 \end{equation}

\begin{equation}
\label{eqn12}
\delta_c =  \frac{1}{3} \frac{\Delta c^3}{(c \: - \: \mathrm{tan}^{-1}(c))} \qquad \mbox{SIS}
 \end{equation}
In the first part of the analysis, we have chosen the virial mass, $M_v$, and the concentration index, $c$  to describe these two-parameter, spherically symmetric density distributions. In our comparison, we use two typical values for the concentration index of the NFW profile: $c=5$, which corresponds to typical cluster mass scale, ($10^{14}\!-\!10^{15}\Msun$), and $c=10$, for cD galaxy scale halos, $\sim 10^{13} \Msun$ \citep{Will00a}. The former, hereafter referred as NFW5, corresponds to the typical lensing group or cluster in our sample while the latter (NFW10) includes the cases where the strong lensing is dominated by the central galaxy alone. This comparison of two concentration indices (at the same overall virial mass) is intended only to quantify the effect of the underlying scale radius on the LOSVD. Furthermore, \citet{Come07} have pointed out that the actual value of \emph{c} in clusters may be $2\sim3$ times higher than the earlier estimates; in the current analysis, therefore, we avoid ambiguity and do not associate the NFW profiles with any names for the objects they may represent. Using the NFW5 profile as the benchmark, the concentration indices for the HRQ and SIS are adjusted to match the density profiles at the scale radius, $r_s$; this follows the approach adopted by \citet{Will00a} in their comparison of the NFW and Kravtsov profiles in weak lensing. The resulting density profiles, scaled by the critical density, are shown in Figure \ref{Fig1} (top, left panel).

\indent With the enclosed mass distribution, Eq \ref{eqn9}, quantities required to compute the LOSVD may be derived. Of these, the radial velocity dispersion, $\sigma_r(r)$, is obtained as the solution of the Jeans equation for hydrostatic equilibrium \citep{Binn80}. Under our assumption of spherical symmetry, we neglect the effect of the anisotropy factor, $\beta = 1 - (\sigma_\theta / \sigma_r)^2$. Our assumption is based on results from dark matter simulations \citep{Cole96} on cluster scales, which have borne out that $\beta \sim 0$, thus lending support to the assumption of isotropy in the radial velocity dispersion. Under this assumption, the scaled radial velocity dispersion may be expressed as,
\begin{equation}
\label{eqn15}
\frac{\sigma^2_r(s)}{V_v^2} =  \frac{1}{\rho(s)} \int_s^\infty \frac{\rho(k)M(k)}{k^2}\: dk
\end{equation}
The circular velocity at the virial radius, $V_v$ is the scaling factor. Using an IDL implementation of the numerical integration routines in \emph{QUADPACK} \citep{Pies83}, we compute the double integral for $\sigma_r(r)$. The scaled circular velocity, $V(s)/V_v$, and radial velocity dispersion, $\sigma^2_r(s)/V_v^2$, for the three density profiles are compared in the bottom panels in Figure \ref{Fig1}.  Our results for the NFW profile are consistent with those of \citet{Loka01} for a similar halo, thus validating our method.  

\begin{figure*}[p] 
  \centering
 \includegraphics[width=5.85in]{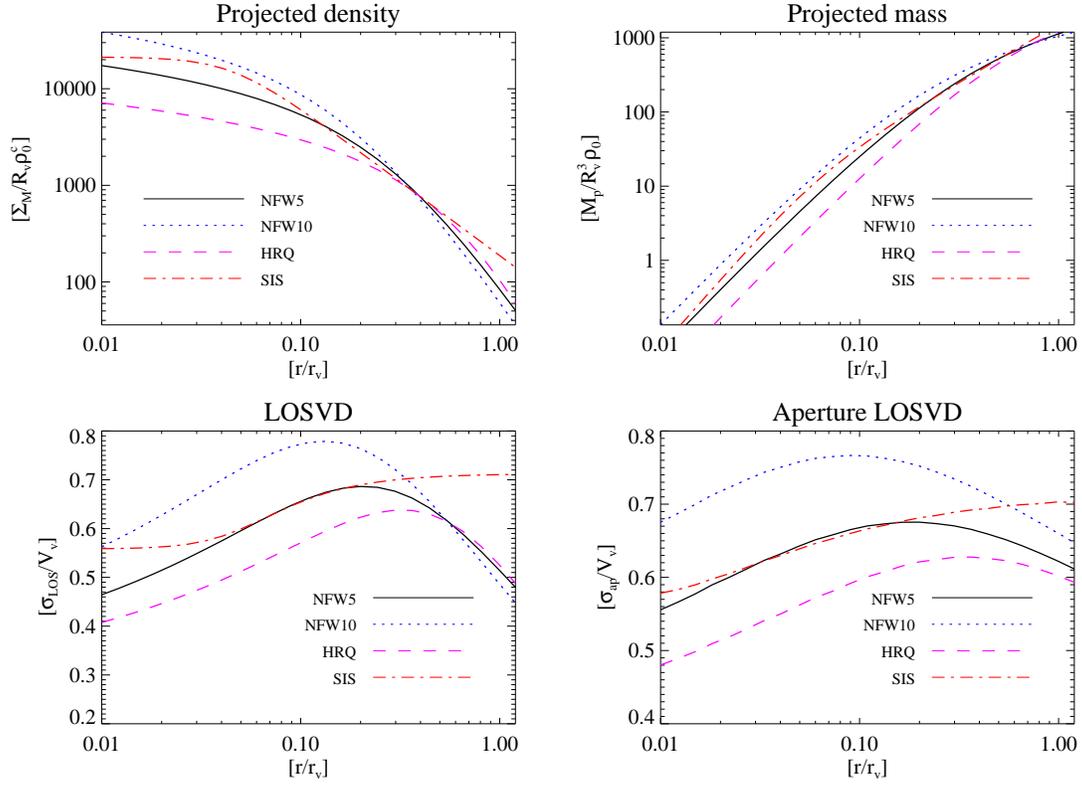} 
  \caption[Comparison of projected properties for NFW, HRQ and SIS density profiles]{Comparison of the scaled \emph{projected} properties of NFW, HRQ and SIS density profiles. The top panels show the scaled surface density, $\Sigma_M/r_v^3\rho_0^3$ (left), and the projected mass, $\Sigma_M/r_v\rho_0^3$ as functions of the scaled \emph{projected} radius, $R/r_v$; the corresponding bottom panels are the LOSVD, $\sigma_{LOS}/V_v$ at an aperture radius, R, and the aperture LOSVD, $\sigma_{Ap}/V_v$}
   \label{Fig2}
\end{figure*}

\indent For our comparison of strong lensing and galaxy dynamics in different dark matter density distributions, we need the projected mass and the LOSVD within an observed aperture, referred explicitly as the \emph{aperture LOSVD} in any context where there may be ambiguity. Expressions for the projected properties of the halo, namely the surface density, $\Sigma_m$, the projected mass, $M_p$ and the LOSVD, $\sigma_{LOS}$, are derived first by integrating the radial profiles of the corresponding halo properties along the line of sight. At an \emph{observed} 2-D aperture radius, R, the surface density is given by,
\begin{equation}
\label{eqn16}
\Sigma_m(R) =  \int\limits_{R}^{\infty} \frac{r\:\rho(r)}{\sqrt{r^2 - R^2}}\: dr
\end{equation}
where, r is the 3-D \emph{intrinsic} radial coordinate. For numerical stability, we carry out the integration with a change to an angular variable, $\psi$ (measured counter-clockwise from the axis along the line of sight), such that the \emph{scaled} surface density may be expressed as,
\begin{equation}
\label{eqn17}
\frac{\Sigma_m(R)}{r_v\rho_0^c} =  \int\limits_{0}^{\pi} R'\:\rho(R') \mathrm{cosec}(\psi)\: d\psi
\end{equation}
where the variable, $R'=R\:\mathrm{cosec}(\psi)$. The projected mass within the observed aperture, R, obtained by integrating the surface density within the aperture, is given in units of $r_v^3\,\rho_0^c$ by, 
\begin{equation}
\label{eqn18}
M_p(R) =  \int\limits_{0}^{R} K\: \Sigma_m(K)\:dK
\end{equation}
with K being a variable of integration. The corresponding LOSVD, $\sigma_{LOS}$, at the 2-D radius, R, is obtained as the density weighted radial velocity dispersion. Using the results from Eq \ref{eqn9}, \ref{eqn15} and \ref{eqn16}, the \emph{scaled} LOSVD may be expressed as,
\begin{equation}
\label{eqn19}
\frac{\sigma^2_{LOS}(R)}{V_v^2} = \frac{1}{\Sigma_m(R)} \int\limits_{0}^{\pi} R'\:\rho(R') \sigma^2_r(R')\:cosec(\psi)\: d\psi
\end{equation}
where, as in Eqn \ref{eqn17}, $R^\prime$ is the scaled radial coordinate. Finally, the aperture LOSVD, $\sigma_{ap}$, is the surface density weighted value of $\sigma_{LOS}$ within the observed aperture. Using Equations \ref{eqn16}-\ref{eqn19}, the aperture LOSVD, in units of $V_v^2$, is obtained from,
\begin{equation}
\label{eqn20}
\sigma^2_{ap}(R) = \frac{2\pi}{M_p(R)} \int\limits_{0}^{R} K\:\Sigma_m(K) \sigma^2_{LOS}(K)\: dK
\end{equation}
Comparative plots of these projected quantities for the different density profiles are shown in Figure \ref{Fig2} and discussed in the following subsection.


\subsection {Results from LOSVD comparison - I \label{LOSres1}}  

We first establish a baseline comparison of the intrinsic and projected characteristics of the three density profiles. Assuming a $10^{15} \msun$ halo and using the NFW5 profile as reference, we adjust the concentration indices of the HRQ and SIS distributions such that their scaled densities match that of the NFW5 profile at the scale radius, $r_s = 0.2\;r_v$ (top left, Figure \ref{Fig1}). The NFW10 profile provides a comparison for variations in the concentration index; the scale radius is this case is $0.1\;r_v$. 

\indent The intrinsic properties of the density distributions as functions of the scaled radius, $s$ are shown in Figure \ref{Fig1}, and their projected counter parts as functions of the scaled \emph{aperture} (= projected) radius, $R/r_v$ in Figure \ref{Fig2}. Since the density distributions have been matched, the principal differences in the intrinsic properties arise mainly in the velocity profiles. The radial velocity dispersion of the centrally concentrated NFW10 shows a higher maximum ($0.8\, V_v$) at a smaller radius, $s= 0.08$, compared to the NFW5 ($0.7\, V_v$ at $s =0.13$) and the Hernquist profiles ($0.66 V_v$ at $s = 0.3$); for the assumed virial mass, the circular velocity at the virial radius, $V_v = 1434 \kms$. In comparison, the velocity dispersion of the SIS model remains uniform at $0.56 V_v$ within the core radius, then rises smoothly to plateau at $0.71 V_v$ at larger radii ($r \gg r_c$). The differences between the radial velocity dispersions, however, are significant only within the core and the profiles, except the SIS, converge toward each other at larger radii (at $r/r_v \geq 0.1$), Even though the radial velocity dispersion is not directly observed, these radial trends affect the projected LOSVD, which is the measured quantity.

\indent The projected properties reflect the cumulative variations between these profiles and show significant differences both in the projected mass as well as in the LOSVD. \emph{The variations in these observed properties are particularly relevant for our work since lensing properties are governed by the projected mass, while the intrinsic galaxy dynamics are probed by the aperture LOSVD}. As expected from the variations seen in the intrinsic properties, the principal differences lie within the core regions and become smaller at larger aperture radii; for instance, at $R/r_v = 0.8$, the observed LOSVD of NFW5 and HRQ are $75\!-\!100 \kms$ below that of NFW10, and $150 \kms$ lower than the SIS at this radius. Therefore, even in the case of matched densities, we expect to distinguish between the SIS and the two other density profiles with the observed LOSVD.  The predicted differences between NFW5 and HRQ for this case however, are smaller than the expected errors in our observations, $\sim 100 \kms$, making it difficult to distinguish between the two; this prediction is consistent with the observational results from the CIRS survey, \citep{Rine06}, discussed earlier in Section \ref{review}. It is for this reason that the \emph{additional constraints provided by strong lensing, which are sensitive to the projected mass distributions at smaller radii, $R/r_v \sim 0.1$, become particularly important}, as discussed in the following subsection. \\


\subsection {Results from Lensing + LOSVD comparison - II \label{LOSres2}}  

\indent The observed lens geometry, along with the measured redshifts of the deflector and source, are used to construct a lens model and thus estimate the underlying \emph{lensing} mass distribution. In our lens model, we assume spherically symmetric mass distribution and that the observed lensed arcs form on the Einstein radius corresponding to the redshifts of the deflector and background source. Under these assumptions, the expression for the Einstein angle, $\theta_E$ \citep{Schn92} may be inverted to obtain the projected mass within the Einstein radius. For the \emph{differential} comparison of density profiles, the lensing mass estimate provides only the normalization in the central region, thus justifying the use of this simple lens model. However, in \S \ref{BaryMass} we quantify possible effects an over- or under-estimate of the lensing mass will have on the LOSVD and thus on the inferred overall mass distribution in the galaxy group.  

\indent For the results given in this subsection, we use deflector and source redshifts of the lensing galaxy group, SL2SJ143000 ($z_d = 0.501, z_s = 1.435$), described in Section \ref{DMlensObs}. The comparisons are carried out for five different values of the virial mass, $10^{14}\,-\,10^{15} \msun$ in equal increments, to cover the typical mass scale from galaxy groups to cluster values; it must be pointed out that constraining the projected mass within the Einstein radius makes the virial mass of the system a free parameter. Given this freedom, for each chosen virial mass, the concentration index, $c$, of each profile has to be iteratively adjusted to match the projected mass, $M_p$, within the Einstein radius, $r_E$, to the mass value obtained from the lens model. 
 
\begin{figure*}[htbp] 
  \centering
 \includegraphics[width=5.85in]{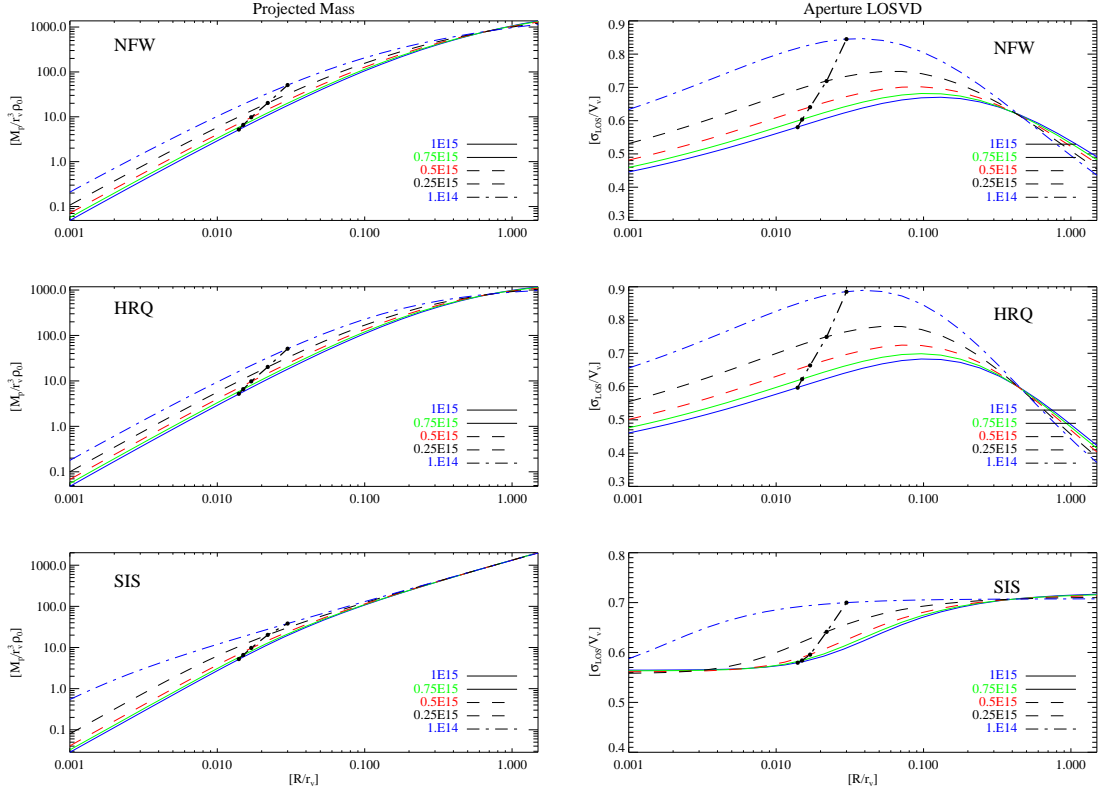} 
  \caption[Effect of virial mass on projected mass and aperture LOSVD]{Effect of the virial mass of the galaxy group or cluster on the projected mass within an aperture (\emph{left panels}), and the aperture LOSVD (\emph{right}) for the NFW, HRQ and SIS density distributions respectively; refer to \S \ref{LOSres2} for model parameters used in these plots. The locus of the projected mass and aperture LOSVD at the \emph{scaled} Einstein radii corresponding to the virial masses is overplotted (chained line)}
   \label{Fig3}
\end{figure*}

\indent Figure \ref{Fig3} shows the resulting projected mass profiles (left panels) and the LOSVD (right) for the NFW, HRQ and SIS profiles as functions of the scaled aperture radius; the effects of increasing virial mass are shown in the five profiles plotted; Table \ref{Tbl1} lists the pertinent numerical values.The projected mass and LOSVD, listed in columns 5 to 10, correspond to a rest frame radius of $0.77 Mpc$ at which the observed LOSVD was measured for SL2SJ143000 (see Section \ref{ObsRes}). 

\indent The effect of \emph{decreasing} virial mass for a fixed projected mass within the Einstein radius, seen in Figure \ref{Fig3}, is to increase the \emph{fraction} of projected mass within the inner radii, as expected; the underlying density distribution becomes more peaked toward the center. The \emph{scaled} aperture LOSVD consequently reflects this central concentration of mass and shows a higher relative peak in each profile as well as an overall increase in the \emph{scaled} LOSVD for decreasing virial mass; however, the increase in the scaled LOSVD is largely offset by the decrease in the circular velocity, (shown in column 3 in Table \ref{Tbl1}), which is used for scaling; therefore, the \emph{observed} value of the LOSVD, in units of $\kms$, is lower for a smaller virial mass.

\indent The peaked trend in the aperture LOSVD for both the NFW and HRQ indicates a significant and \emph{measureable} difference between the value in the core ($r\sim0.2 r_v$) and that at a larger radius, $r\sim r_v$; for example, for the NFW profile, the difference between these two LOSVD values is $\sim 200 \kms$ for a galaxy group scale object. The advantage of using such \emph{differential LOSVD} measurements (measured within a series of apertures) is that they also provide additional measured constraints on the density profile, and thus mitigate any effect of the underestimate of the lensing mass. Given these benefits, differential LOSVD measurements in several apertures may be used to advantage for groups in which the MOS observations confirm a sufficient number of member galaxies needed for the radial binning.


\subsection{ Effect on LOSVD of lensing mass estimate \label{BaryMass}}

In our determination of the lensing mass using a simple lens model, we have made two assumptions that may lead to a corresponding under- or over-estimate of its value. Here, we discuss these assumptions and test what the corresponding effect of such an error would be on LOSVD from our models, and consequently on the concentration index of the density distribution, and virial mass of the group estimated from our observations. First, in our lens model, we assume the deflector mass to be spherically symmetric and the observed lensed images to fall on the Einstein radius corresponding to the lens and source redshifts. The small image separations in group lensing make it reasonable to assume that the source separation from the optic axis is small and therefore the lensed images trace the corresponding Einstein radii. Regarding spherical symmetry, \citet{Cole96} have shown that, on the scale of clusters, the effects of triaxiality in the mass distribution are small, especially in the core of these structures; we assume this to apply to our groups as well. At the same time however, detailed mass models of simulated clusters by \citet{Mene07} indicate that the assumption of spherical symmetry leads to an \emph{underestimation} of 10-35\% in the inferred central slopes and consequently in the total projected mass from lensing; they find that active merging and the presence of substructure leads to the higher degree of underestimation. We therefore introduce underestimates up to 50\% in the lensing mass and investigate the corresponding effect on the velocity dispersion obtained from our model.

\indent Secondly, in the case of lensing groups, the baryonic contribution from the dominant central galaxies to the gravitating mass within the Einstein radius may be significant. In our analysis, we estimate the stellar mass in the central galaxies using the correlation with their luminosities and optical colors, with functional fits provided by \citet{Bell01} and \citet{Bell03}. Systematic errors in this fit would directly give rise to corresponding differences in our estimate of the dark matter within the lensing aperture and thus on the computed velocity dispersion. In this subsection, we assess the effects and the relative importance of these competing effects of under- or over-estimated dark matter within the Einstein radius, $r_E$, on the computed LOSVD at larger radii (up to the virial radius), and therefore on the estimated virial masses from our observations.

\indent To test the effects of variations in the amount of dark matter,  we have generated a set of LOSVD profiles using mass models in which the \emph{`true'} projected dark matter mass within $r_E$ ranges from $50\%$ to $150\%$ of the gravitating mass \emph{estimated} from the lens model; as in the previous section, we use the lensing geometry and parameters of SL2SJ143000, described in \S \ref{DMlensObs}, for this comparison.    

\begin{figure*}[htbp] 
  \centering
  \includegraphics[width=5.85in]{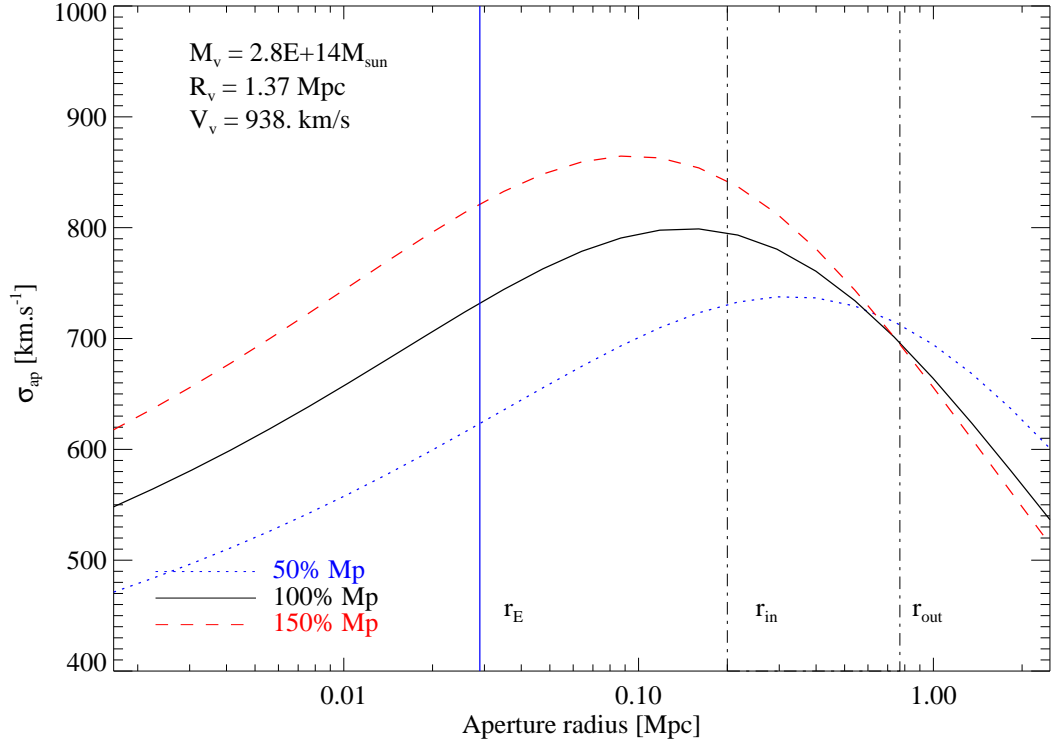} 
  \caption[Effect of under- or over-estimated dark matter lensing mass]{LOSVD profiles reflecting the effect of an under- or over-estimate of the dark matter contribution to the lensing mass. The Einstein radius of SL2SJ143000 and the aperture radii at which the LOSVD for this system have been measured are annotated for reference.}
   \label{Fig4}
\end{figure*}

\indent Figure \ref{Fig4} shows the LOSVD profiles corresponding to a $50\%$ under-estimate of the projected dark matter within $r_E$, as well as a $50\%$ over-estimate, with both these profiles  compared with that of a model in which the lensing mass is $100\%$ dark matter only (our reference). Only the LOSVD profiles corresponding to the NFW density distribution are shown since the effects on the other profiles are similar. For all models, the virial mass of the group is kept fixed at $2.8\times 10^{14}\,\Msun$, as estimated for this group (see details in \S \ref{Modfit}); consequently, the concentration index of the profile is varied to account for the the variations in the projected dark matter mass within the Einstein radius. 

\indent Qualitatively, an over-estimate in the projected dark matter mass within $r_E$ leads to a more centrally concentrated, `peaked' profile with a higher concentration index; this peaked density profile translates directly into the LOSVD profile as well, with the maximum velocity being higher ($\sim10\%$) and occuring at a smaller radius ($\sim10\%$); at the lower end of the scale, the $50\%$ under-estimate has the opposite effect on the LOSVD profile with a lower maximum occuring at a larger radius.

\begin{figure*}[htbp] 
  \centering
 \includegraphics[width=5.85in]{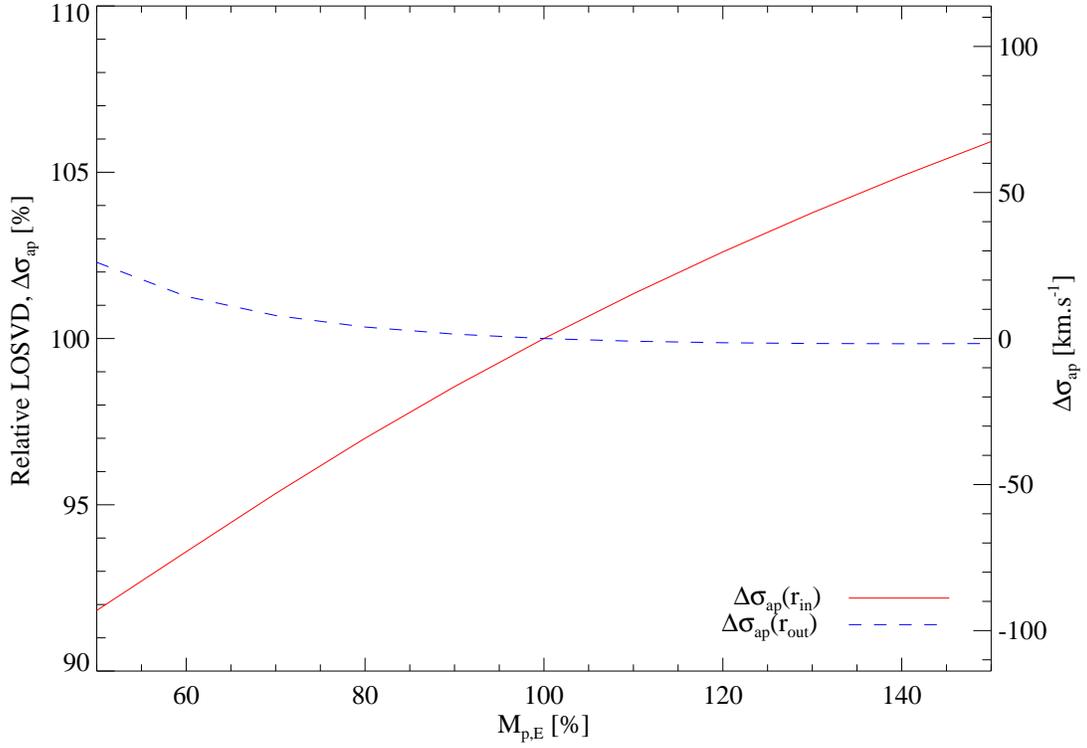} 
  \caption[Percentage change in LOSVD at measured apertures]{Relative change in the aperture LOSVD due to an under- or over-estimation of the dark matter contribution to the lensing mass (expressed as a percentage). The percentage changes in LOSVD are expressed relative to the value corresponding to a dark matter contribution equal to $100\%$ of the lensing mass; the left axes expresses the relative change in percentage, while the right axes shows it in $\kms$. The measured lens geometry and LOSVD values of SL2SJ143000 have been used in this comparison.}
   \label{Fig5}
\end{figure*}

\indent Figure \ref{Fig5} quantifies these variations in the LOSVD due to under- or over-estimates of the dark matter lensing mass; since the quantities of interest  are the LOSVD values at the chosen (observed) aperture radii, the plots trace the variations in LOSVD at the two observed apertures in SL2SJ143000, our chosen example. The abscissa shows the percentage variation in the dark matter mass. The ordinate traces the corresponding percentage variations in the  LOSVD values, normalized using the value for the $100\%$ dark matter model, our reference; also shown on the right axis is the relative change expressed in units of $\kms$, which may be conveniently compared against the expected observational uncertainties.

\indent The trends again show that variations in the dark matter within $r_E$ primarily affect the LOSVD toward the core of the group, and have only a minimal effect on the velocity dispersion measured at larger radii. Even in the core, the LOSVD variation is only $\sim5\%$ for a $50\%$ change in the projected dark matter mass, with the corresponding measured velocity variations being $\sim50\kms$, which is of the order of magnitude of the observational errors (see \S \ref{ObsRes}). Therefore, the assumption of spherical symmetry and any systematic errors in the estimation of the baryonic contribution on the LOSVD are minimal and do not affect the inferences drawn from our \emph{differential comparison of density profiles}. However, these results provide a measure of the error in our inferences of the \emph{absolute} values of the underlying mass distribution. 


\section {Observations} \label{DMlensObs}

The results reported here are from our Gemini GMOS-MOS observations of two lensing groups, SL2SJ$143000+554648$ and SL2SJ$143139+553323$, drawn from a sample of nine candidates available in our lens catalog prepared from 161 $\sqdeg$ of CFHTLS-Wide\footnote{Canada-France-Hawaii Telescope, Legacy Survey,  http://www.cfht.hawaii.edu/Science/CFHLS/} imaging. All of the lensing systems were discovered by visual inspection of RGB color images of candidate groups and clusters detected using an automated cluster detection algorithm, \emph{K2}; \citet{Than09} provide a description of the algorithm, completion and contamination statistics of the detector, as well as a description of the resulting catalogs.

\indent From the MOS observations of the group members, we measure the LOSVD as a proxy for the dynamical masses of the groups. The candidate member galaxies for this MOS follow-up are identified using their optical $gri$ colors and positions relative to the central galaxies. The identification of members is completed by \emph{K2} as part of the group and cluster detection process; see \citet{Than09} for details. Results from our MOS observations are discussed in this section after a brief summary of the instrument configuration and data reduction. 


\subsection{Instrument configuration and target selection \label{InstConf}}

The multi-object spectrograph, Gemini GMOS\footnote{http://www.gemini.edu/sciops/instruments/gmos/} with a FOV of $5'.5\times5'.5$, permits simultaneous observations of several galaxies ($\sim 20$ or more) spanning a sky aperture of a few arc minutes. By choosing target galaxies brighter than $i = 22$mag and using on-chip spatial and spectral binning, with $\sim 2$h integration times we achieve adequate S/N $\geq 5$ per spectral pixel in the region of the strong Balmer break at $4000\angs$ in these early type galaxies, which we use for redshift measurement. The $1\asec$ MOS slitlets and the R400 grating combination provide a redshift sampling, $\Delta z = 0.001$. Each field  is observed with two MOS masks in order to increase the number of candidate member galaxies observed, especially in the crowded central regions of these lensing groups; the integration time per mask is 7200s, with observations completed in August 2007 as part of the Gemini program, GN-2007A-Q-92 (PI.~ K.~Thanjavur). 


\subsection {Data reduction and analysis \label{DataRedn}}

In processing the MOS observations using the standard Gemini-IRAF pipeline, we have improved the fringe correction and sky subtraction by incorporating B-spline based scripts, which have been shown to be advantageous \citep{Kels03}. The extracted 1-D spectra are co-added to make the final spectrum for each observed object. To determine the redshift, the co-added spectra are continuum subtracted and then cross correlated against template galaxy spectra; for the cross correlation, performed in Fourier space, we use the IRAF\footnote{IRAF is distributed by the National Optical Astronomy Observatory, which is operated by the Association of Universities for Research in Astronomy (AURA) under cooperative agreement with the National Science Foundation.} task \emph{FXcor}. For cross correlation, we use representative template spectra for a galactic bulge, Elliptical and S0 galaxies from the \emph{Kinney} catalog \citep{Kinn96, Mcqu95}, with the best fit being determined by the highest \emph{R}-value \citep{Tonr79} of the correlation. 

 
\subsection{Observational results \label{ObsRes}}

\begin{figure}
 \plottwo{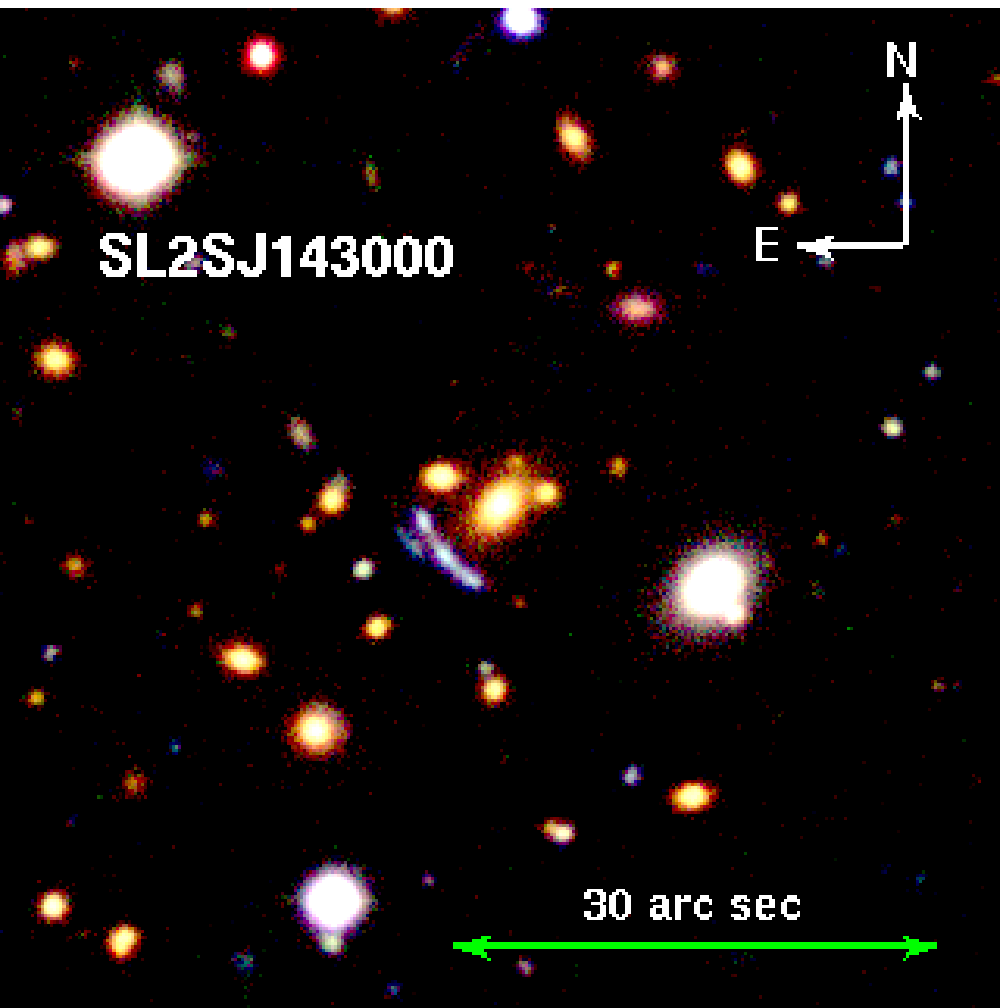}{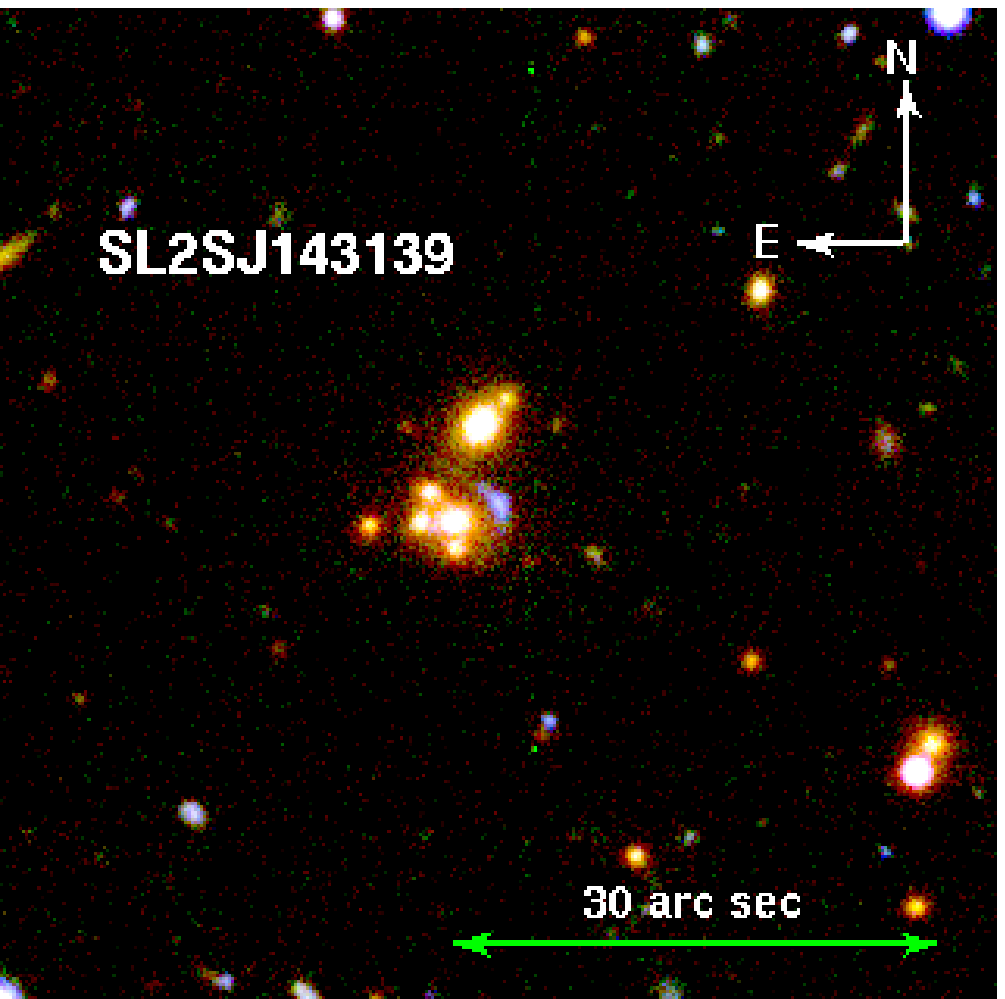}
\caption{RGB color images of a $1'\times1'$ field centered on the two lensing groups, SL2SJ143000 (\emph{left}), and SL2SJ143139 (\emph{right}), discussed in this paper. Both were observed as part of our observational program, Gemini North Telescope, GN-2007A-Q-92. The color images were made from available CFHTLS-W g,r,i imaging of the fields, and are shown in the standard North up, East left orientation. \label{Fig6}}
\end{figure}

\noindent \textbf{SL2SJ143000+554648} (RA 14:30:00.7, Dec 55:46:47.9):  With a classic lensing geometry seen in the RGB color image, Figure \ref{Fig6} (\emph{left}), a background galaxy at a redshift of $1.435\pm0.001$ is lensed by the mass in the central regions of a galaxy group at a redshift of $0.501\pm0.001$. The redshift of the source was determined by longslit observations as part of an earlier Gemini program (GN-2007A-Q-114, PI. K. Thanjavur); the redshift of the group was measured from the GMOS observations described here. The color image shows that the light in the core region of this group is dominated by the BCG, which may be assumed to contribute the majority of the mass in the region and therefore is the principal deflector. Of the 39 objects selected for observation using two MOS masks, 38 matched the early type templates during cross correlation and yielded secure redshifts. Of these, 19 galaxies are consistent within $ \Delta z=\pm0.005$ of the redshift of the BCG at $z=0.497$, and are taken to be members of this group. The bi-weight mean redshift of these member galaxies is taken to be the redshift of the group. The distribution of the galaxies on the sky indicates that 10 members lie within a \emph{projected} aperture radius corresponding to rest frame 0.5 Mpc, shown by the middle circle in Figure \ref{Fig7}; the small and large apertures are used for LOSVD comparisons and are described in Section \ref{LOSres3}. Table \ref{Tbl2} lists the spectroscopic redshifts, measured photometry and the sky positions of the member galaxies relative to the BCG. Figure \ref{Fig8} (left panel) shows the velocity distribution of the member galaxies as well as the fit to the LOSVD, both of which are discussed further in Section \ref{LOSres3}.  

\noindent \textbf{SL2SJ143139+553323} (RA 14:31:39.00, Dec 55:33:23.00): In this interesting lensing group, shown in the RGB color composite in Figure \ref{Fig6} (\emph{right}),  four galaxies with colors consistent with early type galaxies are clustered within a \emph{projected} aperture of radius $\sim20 \kpc$ in the rest frame of this group at $z=0.669\pm0.002$. The blue lensed image, lying to the NE of these central galaxies, appears distorted by the complex mass distribution. The chosen field orientation during observation permitted slitlet placement on 25 objects, which yielded secure redshifts for 22 early type galaxies, of which 9 are consistent with the median redshift of the group. The projected radial distribution, Figure \ref{Fig9} shows that 7 of these members lie within a projected distance of 0.5 Mpc of the center of the group, the barycenter being taken to coincide with the center of the light distribution of the four galaxies in the core. Measured values of spectroscopic redshifts, photometry and sky positions of the member galaxies are listed in Table \ref{Tbl2}; their velocity distribution and the LOSVD fit are shown in Figure \ref{Fig8} (\emph{right}).  

\indent For the lensed arc, the recent release of \emph{Terapix} T0006 (Nov 2009) photometric catalogs containing $u^*$- and $z$-bands magnitudes for this CFHTLS-W field, along with the $gri$ photometry available earlier, permit us to determine the photometric redshift. The arc is clearly detected in all five filters. Using $HyperZ$ \citep{Bolz00} with this 5-filter data, we determine the photometric redshift of the arc as $z_s=2.083\pm0.07$. The $HyperZ$ probability of the fit is $> 60\%$ for the best fitting Sbc galaxy template; other late-type galaxy templates yielded similar redshifts.  In addition, during our GMOS-MOS observations, we devoted a slitlet to the lensed arc but were unable to determine the redshift due to the lack of any emission line; the absence of an emission line in the wavelength range (up to $1\mu m$) covered during our spectroscopy places the source at a redshift $>1.5$, consistent with the $HyperZ$ redshift estimate. HST WFPC-2 imaging of SL2SJ143139, observed as part of a Cycle 16 Snapshot program (PI. J-P. Kneib) awarded to our SL2S collaboration, shows the distinctive features of a bone fide lensed arc, thus ruling out the possibility of a foreground edge-on galaxy; based on the available data, modeling of this lens system by our SL2S collaborators is in progress.
 

\subsection{Rest frame LOSVD estimates \label{LOSres3}}

For the computation of the LOSVD from the measured redshifts of the member galaxies, we adopt the \emph{bi-weight} estimator in the public statistical package, \emph{ROSTAT} \citep{Timo90}. Using extensive simulations, \citet{Timo90} have shown that the \emph{bi-weight} is resistant to errors due to small number statistics (our samples typically contain only 10 to 20 galaxies with secure redshifts), and to be especially immune to outliers; galaxy groups are prone to outliers since \emph{individual} member galaxies may be in the initial phases of infall or may be undergoing interactions among themselves; see discussions in \citet{Gebh91} and \citet{Beer91}. 

\indent The estimated LOSVD for the two galaxy groups, corrected to their rest frame values, along with their confidence intervals are listed in Table \ref{Tbl3}. The error in the redshift of each galaxy from the cross correlation is taken into account by  \emph{ROSTAT} during the estimation of the \emph{bi-weight}. The $1\sigma$ confidence intervals on the spread, given in the table, are  estimated in  \emph{ROSTAT} with a bootstrap technique with 1000 resamplings (see \citet{Timo90} for details on the methodology). 

\begin{figure*}[htbp] 
  \centering
 \includegraphics[width=5.85in]{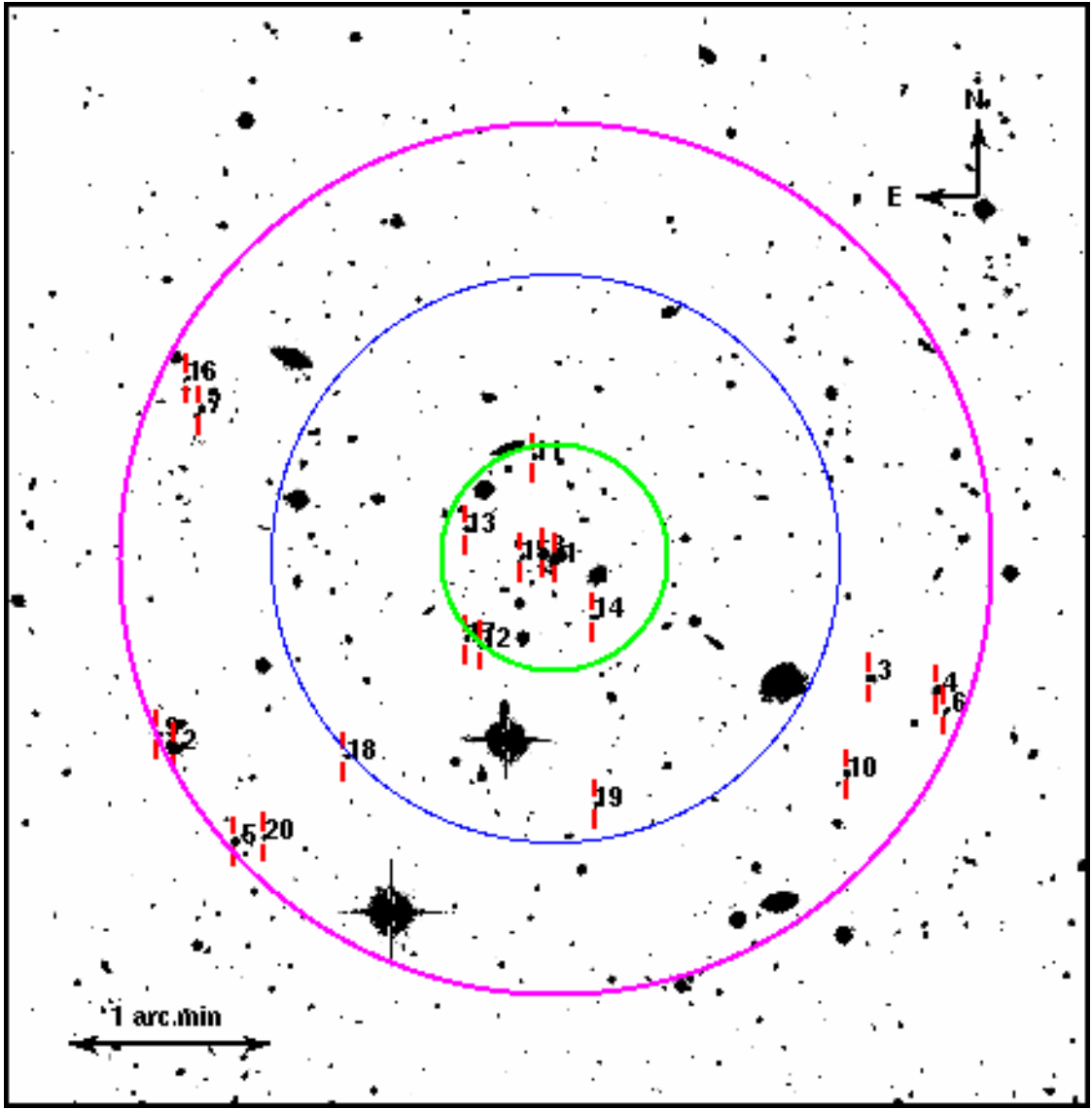} 
  \caption[Observed cluster members of SL2SJ143000]{A $5'.5\times5'.5$ CFHTLS-W i'-image showing the observed GMOS-MOS field centered on SL2SJ143000 at z=0.501, with the confirmed cluster members indicated (numbered in descending order of brightness). An aperture corresponding to $0.5\mpc$ at the redshift of the cluster is shown for reference (middle aperture). Also shown are the inner ($0.2\mpc$) and outer ($0.77\mpc$) apertures used for LOSVD measurement (see Section \ref{LOSres3}}
   \label{Fig7}
\end{figure*}

\begin{figure}
 \plottwo{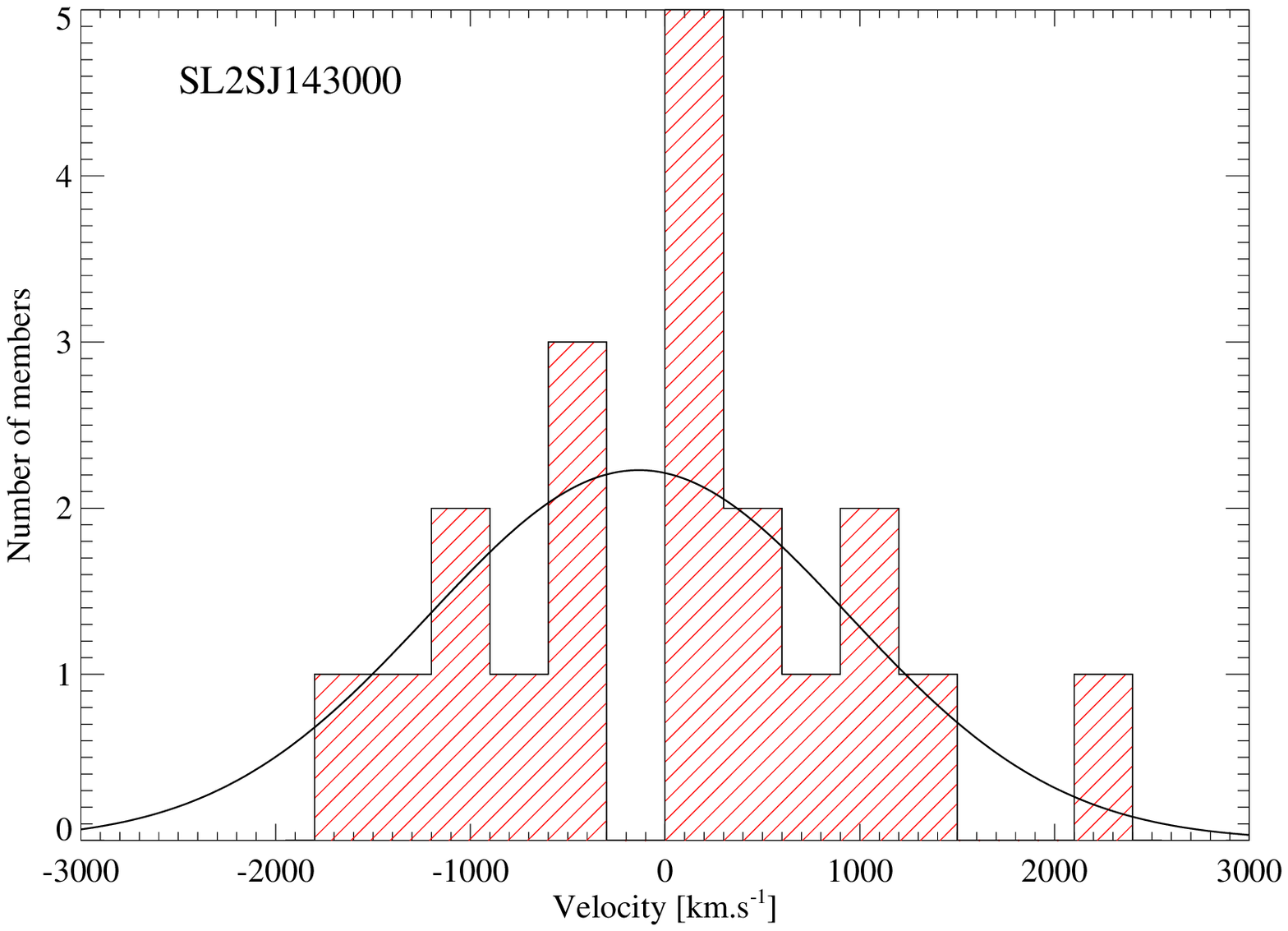}{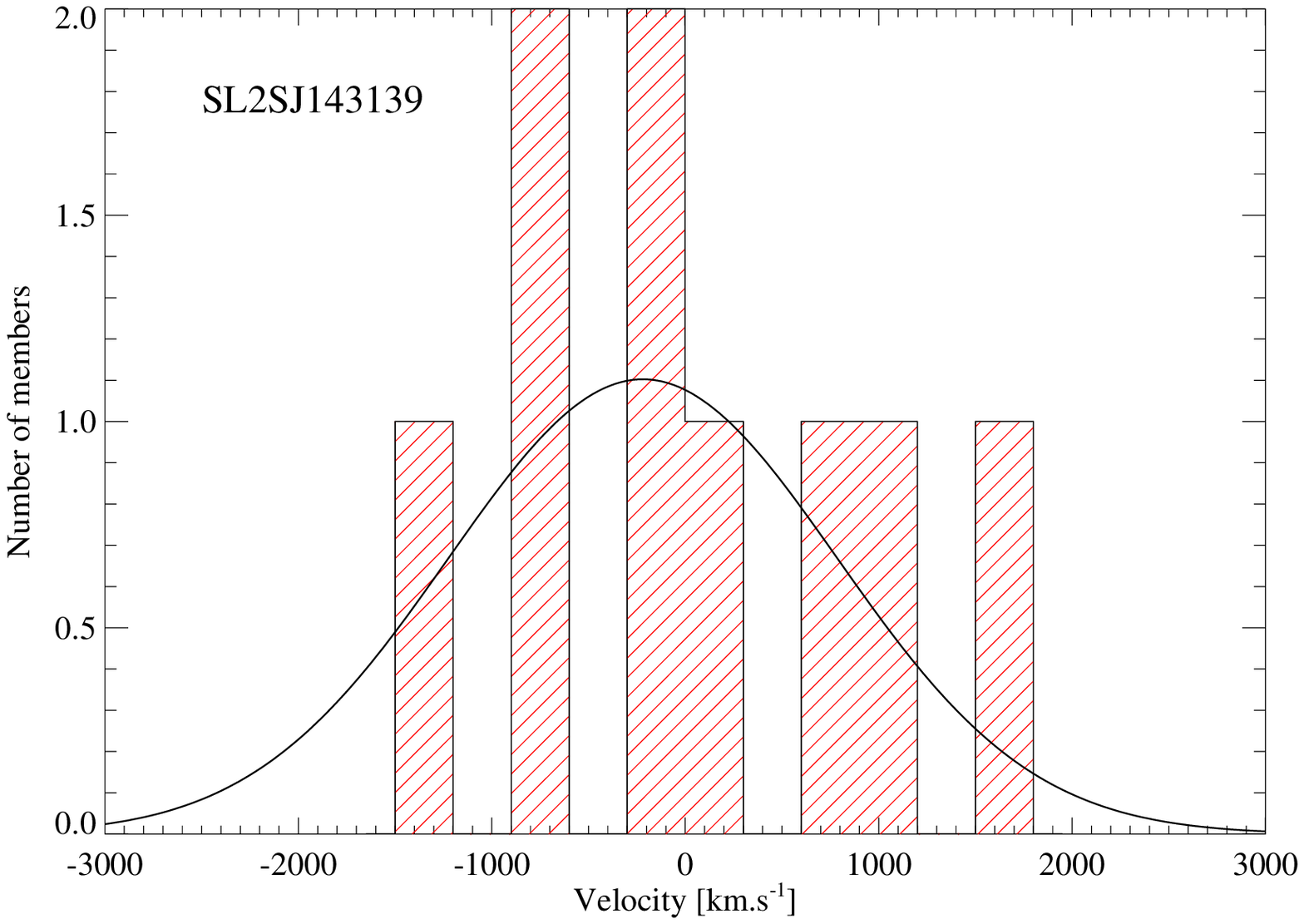}
\caption{ Histograms showing the velocity distributions of the member galaxies of SL2SJ143000 (\emph{left}) and SL2SJ143139 (\emph{right}) about the bi-weight mean redshifts of each group. Overplotted are representative Gaussians generated using the parameters estimated by \emph{ROSTAT}, as described in Section \ref{LOSres3}
\label{Fig8}}
\end{figure}

\begin{figure*}[htbp] 
   \centering
 \includegraphics[width=5.85in]{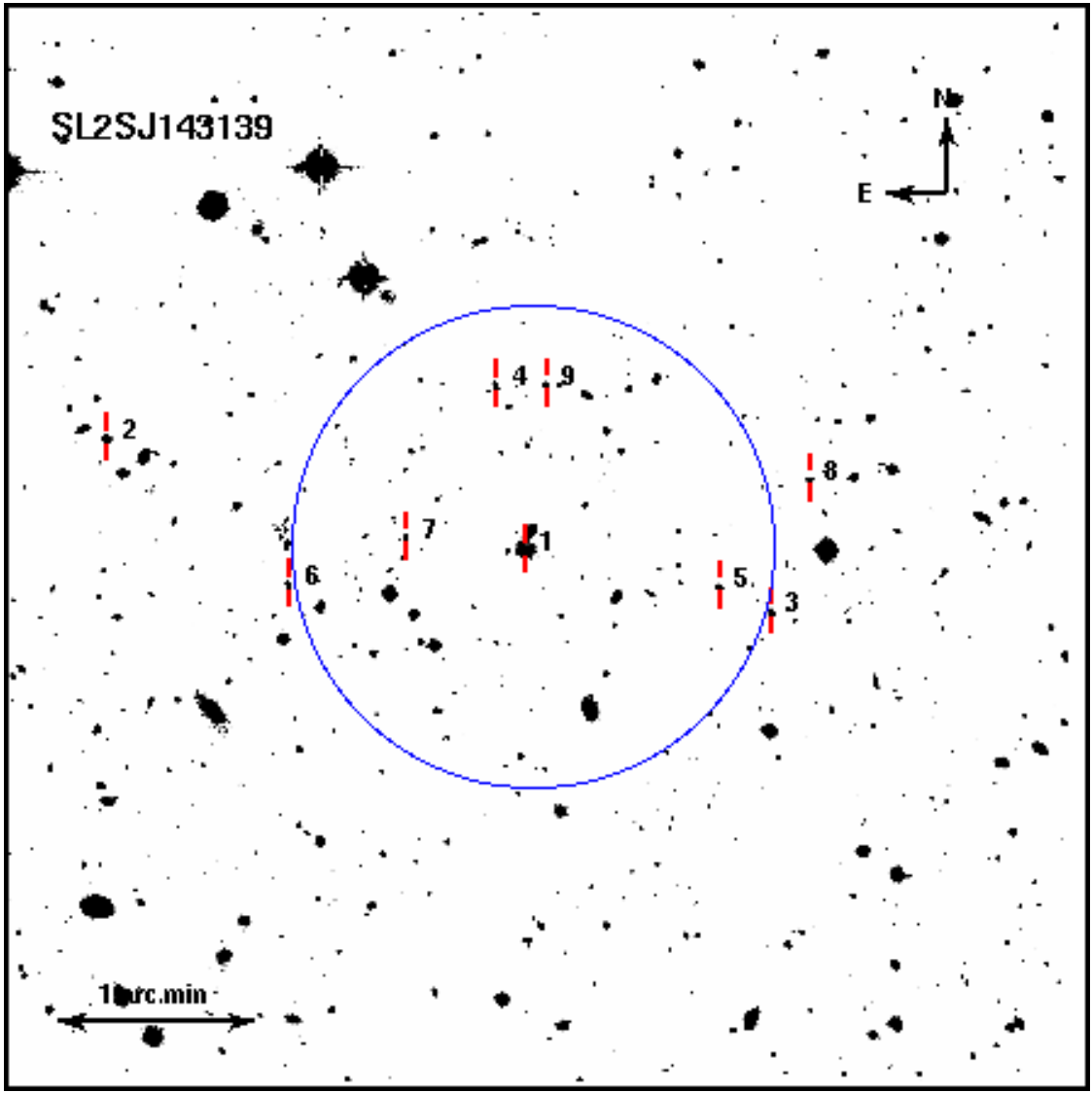} 
   \caption[Observed cluster members of SL2SJ143139]{A $5'.5\times5'.5$ CFHTLS-W i'-image showing the observed GMOS-MOS field centered on SL2SJ143139 at z=0.669, with the confirmed cluster members indicated (numbered in descending order of brightness). An aperture corresponding to $0.5\mpc$ at the redshift of the cluster is shown for reference.}
    \label{Fig9}
 \end{figure*}

\indent Figure \ref{Fig8} (\emph{left}) is a histogram of the measured velocity distribution of the 20 member galaxies of SL2SJ143000 about the mean redshift of the group, $z = 0.501\pm0.001$; overplotted is a representative Gaussian distribution generated using the parameters estimated with \emph{ROSTAT}. The corresponding rest frame LOSVD is $720^{+91}_{-110}\kms$ within an aperture of radius $0.77 \mpc$ at the redshift of the group (shown by the large aperture in Figure \ref{Fig7}).   

\indent Figure \ref{Fig8} (\emph{right}) shows the velocity distribution of the 9 observed cluster members of SL2SJ143139 about the group redshift, $z = 0.669\pm0.002$; the estimated rest frame LOSVD is  $563^{+87}_{-137} \kms$ within an aperture of $0.912 \mpc$. 

\indent For SL2SJ143000, we measure a LOSVD of  $784^{+126}_{-115} \kms$ for a subset of 6 members within a rest frame aperture of $0.2 \mpc$ (shown by the small aperture in Figure \ref{Fig7}); this is the minimum number of galaxies needed for meaningful LOSVD measurement, according to results from  \emph{ROSTAT} simulations \citep{Timo90}. This second measurement not only provides an additional constraint for the estimation of the virial mass, but also increases the sensitivity of our method to distinguish between the underlying density distributions. Due to the peaked profile of the aperture LOSVD as a function of aperture radius, seen in Figure \ref{Fig2}, we expect observable differences in LOSVD values between the NFW and the Hernquist profiles at smaller radii ($r/r_{vir}\sim0.2$).

\indent In the following subsection, we apply our formalism presented in \S \ref{LOScomp} to convert the observed LOSVD and estimated lensing mass (using the lens model described in \S \ref{LOSres2}) into the corresponding virial mass and M/L ratio of each group. 


\subsection{Estimation of the viral mass and M/L ratio \label{Modfit}}

Using a maximum likelihood parameter estimator in conjunction with the formalism presented in \S \ref{LOScomp}, we determine the virial mass of each group and the corresponding concentration index for each of the NFW, HRQ and SIS density distributions applying constraints from the measured dynamical and lensing masses. With the summed $i$-band luminosity within the virial radius, we then estimate the M/L ratio of these structures.

\indent In our models, the LOSVD estimates determined in \S \ref{LOSres3} are direct proxies for the dynamical masses enclosed within the corresponding apertures. For estimating the lensing mass, we use the lens model described in \S \ref{LOSres2}. The Einstein angle, $\theta_E$, needed for the estimate is measured from available CFHTLS $g$-band imaging. The source redshifts are known from the longslit or MOS spectroscopy carried out to confirm these systems are bone fide lenses; the redshift of the group, which is the deflector, is known from the MOS spectroscopy described here. For the lensed source in SL2SJ143139, we have used available 5-filter CFHTLS-W photometry to estimate the photometric redshift with a high degree of confidence; near infrared spectroscopy of this object is planned. In must be emphasized that increasing the redshift to even z=3, alters the computed lensing mass by $<10\%$, while the consequent uncertainty introduced in the virial mass is $\ll5\%$. Results for this system presented here will therefore see a revision of \emph{at most} a few percent when we confirm the source redshift with our near-infrared spectroscopy.

\indent From the estimated \emph{total} lensing mass, we subtract the baryonic contribution of the central galaxies within the Einstein radius to obtain the projected dark matter mass; this is the observable we use in our models to trace the underlying density distribution. In estimating the baryonic mass, we assume it to be composed entirely of the stellar content of the galaxies and neglect any contribution from hot gas. Using the i-band luminosities and (r-i) colors of the central galaxies, taken from $Terapix$ photometric catalogs\footnote{Available from the Canadian Astronomy Data Centre (CADC)} for these CFHTLS-Wide fields, we compute their $i$-band stellar M/L ratios and thus the stellar masses using the correlation proposed by \citet{Bell01},
\begin{equation}
\textrm{log}(M/L)_i\ =\ 0.006\ +\ 1.114(r\;-\;i)
\label{clr2ML}
\end{equation}
The correlation coefficients for the SDSS filters are taken from \citet{Bell03}, (Table 7 in their Appendix 2); the solar i-band absolute magnitude, $M_{i,\odot}\,=\,4.48$ is used to convert the M/L ratio to stellar mass in that filter. Table \ref{Tbl4} lists the lensing details of each group, the estimated total lensing mass as well as the baryonic and dark matter contributions, along with the estimated uncertainties in all these estimates. The uncertainty in the total lensing mass is governed by the redshift errors while the error in the baryonic mass estimate depends primarily on the associated photometric errors, obtained from the \emph{Terapix} catalogs.  
 
\indent The maximum likelihood estimator accepts the LOSVD and the corresponding aperture radii, and the lensing dark matter mass and the measured Einstein radius as constraints in fitting for the virial mass, $M_v$ and the concentration index, $c$, using the Marquardt-Levenberg algorithm \citep{Pres92}. The $\chi^2$ of the goodness of fit is computed as,
\begin{eqnarray}
\chi^2\ &=& \left(\frac{M_{p,obs}-M_{p,mod}}{\sigma_{M_p}}\right)^2 \nonumber \\
&+&\;\sum^n_{i=1}\left(\frac{LOS_{obs}(r_i) - LOS_{mod}(r_i)}{\sigma_{LOS(r_i)}}\right)^2 
\label{fitchi2}
\end{eqnarray}
where the subscripts on the projected lensing mass, $M_p$, and the LOSVD refer to the observed and the modeled values; the uncertainty in each quantity is indicated by $\sigma$ with the appropriate subscript. The estimation of the LOSVD within multiple apertures, as in the case of SL2SJ143000, provides additional constraints with consequent improvements in the reduced $\chi^2$ value of the fit; the summation in Eqn \ref{fitchi2} takes this case of multiple apertures, ($r_i$, i=1..n), into account. For SL2SJ143000, n=2, while for SL2SJ143139, n=1. During the iterations for the virial mass and the concentration index, the corresponding model values of the projected mass and LOSVD are computed using Eqn \ref{eqn18} and Eqn \ref{eqn20} respectively. The estimator is applied for each of the assumed density profiles independently and the corresponding best fit parameter values and the associated uncertainties are listed in Table \ref{Tbl5}.

\begin{figure*}[htbp] 
   \centering
 \includegraphics[width=5.85in]{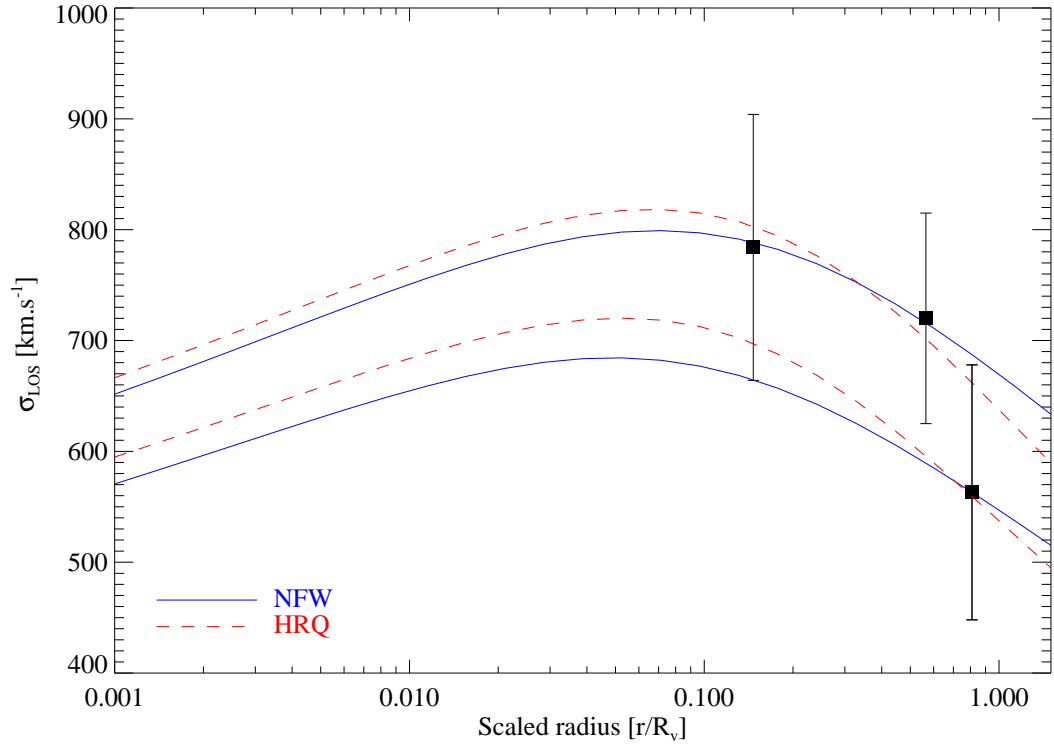} 
   \caption[Comparison of observed and predicted LOSVD]{Comparison of the measured LOSVD values for the two lensing groups from our GMOS observations against the predicted velocity profiles corresponding to NFW and HRQ density distributions. The SIS distribution is rejected at $\gg3\sigma$ by the likelihood estimator and is therefore not shown in the figure. The error bars for the observational data represent bootstrapped $1\sigma$ confidence intervals from \emph{ROSTAT}.}
    \label{Fig10}
 \end{figure*}

\indent For the NFW and HRQ profiles the parameter estimator converged for acceptable values of the reduced $\chi^2$; for the SIS however, the convergence was very poor with the reduced $\chi^2$ exceeding ten for both groups. A comparison of the measured LOSVD and lensing mass for each group, used as the input values for parameter estimation, and the corresponding computed values using the best fit virial mass and concentration index for each density profile is shown by the values listed in Table \ref{Tbl6}. The complementary Figure \ref{Fig10} shows a comparison of the computed LOSVD profiles, using the best fit parameters for the NFW and HRQ density distributions, against the observed values for the two groups; given the poor convergence obtained for the SIS distribution, the corresponding LOSVD is not shown. The plots indicate that the modeled LOSVD estimates for either density distribution are consistent with the observations, well within the associated uncertainties. This compatibility with either density distribution was reported by \citet{Rine06} for their sample of 72 low redshift clusters. Our ongoing observations of the remaining candidates in our lens catalog will determine if this applies to group scale structures as well. 

\indent To compute the $i$-band M/L ratio, we sum up the total $i$-band luminosity of the candidate member galaxies within the virial radius, $r_v$, corresponding to the estimated virial mass. Member galaxies are identified using a color cut of $\pm0.25$mag relative to the ($g-r$) and ($r-i$) colors of the central galaxies. Based on the associated photometric errors, only galaxies brighter than $i \leq 22.5\mag$ are used; this magnitude limit is $\sim2\mag$ fainter than a typical $M^*$ galaxy at z=0.5. The width of the color cut is based on the color errors of the spectroscopically identified cluster members; tests carried out on our cluster and group detection algorithm, $K2$, \citep{Than09} have shown this multi-color based approach of identifying member galaxies to be effective. Table \ref{Tbl7} lists the number of members identified for each group and their total luminosities and stellar masses. Using these values, the total M/L ratio for each group for an underlying NFW and HRQ density distribution and the associated uncertainty are computed and listed in Table \ref{Tbl7}; given the poor fit for the virial mass obtained for the SIS density distribution, we do not compute the corresponding M/L ratio.


\subsection{Comparison with previous estimates \label{CompRes}}

\indent Here we compare our results with those of previous investigations regarding the assembly of mass at the scale of galaxy groups. Pertinently, the joint strong and weak lensing analysis on the SL2S sample by \citet{Limo09} included one of our groups, SL2SJ143000, and thus provides an independent verification of our methods and results. 

\indent For SL2SJ143000, we \emph{measure} an Einstein radius of $4\asec.6$ on the $i$-band image, which closely matches the value of $4\asec.69$ which \citet{Limo09} derive for this system using their lens model. This close match lends support to our assumption that in these lensing groups with image separations of a few arc seconds, the position of the lensed image coincides closely with that of the Einstein radius corresponding to the deflector and source redshifts, The weak lensing analysis of \citet{Limo09} yields an upper bound for the LOSVD of $684\kms$ compared to $720^{+91}_{-110}\kms$ we determine from our MOS spectroscopy, compatible well within the uncertainties of these two estimates. Our estimate of the virial masses corresponding to the NFW and HRQ density profiles are lower than their weak lensing \emph{upper bound} estimate of $6.8\times 10^{14}\msun$. It should be noted that they assume an SIS profile which our models discount. Even with this caveat, our $i$-band M/L ratio estimates for the NFW and HRQ profiles match their value of $\sim250$ also in the $i$-band for the sample of 13 groups. We are extending our analysis to other groups in our catalog and will report in a subsequent publication further comparisons between these two approaches. 

\indent \citet{McKe09} combine strong lensing and galaxy kinematics, similar to our approach, for their recent analysis of the gravitational lens, CLASS B2108+213 at $z_d=0.463$. The observed velocity dispersion is $694\pm94\kms$ for the 51 confirmed galaxy members, though the authors interpret the observed bimodality in the distribution of the velocities as being due to two merging groups. A detailed lens model, based on HST and Keck UBVRI imaging, along with the measured dynamics indicates a density profile which is \emph{isothermal} within the Einstein radius. Even though the estimated virial mass of the group is not reported, the authors conclude that the observed velocity dispersion, which is higher than the range expected in low redshift groups, indicates that two groups may be in the process of merging to form a poor cluster. The condensation of baryons to the core of the merging structure may be the cause of the higher slope of the density distribution deduced by their lens model.

\indent Using stacked weak lensing analysis of 116 groups from the CNOC2 survey, \citet{Park05} report a mean $B$-band M/L of  $185\pm28$ for the complete sample; in addition, they split their sample into `poor' and `rich' categories, and determine M/L$=278\pm42$ for the rich category - a value which indicates that both our lensing groups fall in this class of groups. However, the \emph{singular isothermal} velocity dispersion, $\sigma_{SIS}=270\pm39\kms$ for their `rich' groups is $<50\%$ of the values we measure for both groups from our observation. \citet{Limo09} point out that the M/L values plateau off for structures with virial masses, $M_v\geq10^{14}\msun$ while the velocity dispersions continue to increase from $\sigma\sim300\kms$ for group scale structures to $\sigma\sim1000\kms$ observed in clusters \citep{Bard07}. 

\indent \citet{Shel09b} present the M/L values of their SDSS DR4-MaxBCG cluster sample binning them by richness, $N_{200}$, as well as by their $i$-band optical luminosity, $L_{200}$, thus permitting us to compare our M/L results in each category separately. Assuming that the number of member galaxies in each of our groups listed in Table \ref{Tbl7} is equal to their tabulated $N_{200}$ value ( \citet{Shel09b}, Table 1), the corresponding $i$-band M/L for the MaxBCG sample is $\sim25\%$ \emph{higher} than the values we obtain from our analysis; eg. the MaxBCG M/L = 319 $\msun/\Lsun$ for a SL2SJ143000-like group with $N_{200} = 20$, compared to the M/L value of 265 we estimate for this group. The MaxBCG members are identified using color cuts with reference to the Red Sequence, similar to the method used by our cluster detector, $K2$, as described in \S\ref{Modfit}. However, the $N_{200}$ values used for comparison must be corrected for differences in the magnitude limits used for cluster member selection as well as the median redshifts of the two cluster catalogs. Using the MaxBCG selection criteria from \citet{Koes07}, and the $K2$ limits for cluster members given in \S \ref{Modfit}, we compute the correction. For this we use the SDSS $i$-band Schechter parameters at z = 0.1 from \citet{Blan03} and compute the number density for each case; we assume that there is negligible evolution in the Schechter parameters up to z=0.5, the nominal redshift of our galaxy groups. Applying this correction reduces the equivalent MaxBCG $N_{200}$ value for a SL2SJ143000-like group to 12 with a corresponding M/L $\sim250\msun/\Lsun$, thus making our estimates consistent. Similar consistency is also obtained in the corrected M/L values of both samples determined using their $L_{200}$ values.

\indent These comparisons indicate that the estimated M/L ratios of our candidates are consistent with other published results, and represent rich group scale objects. Their virial masses and observed LOSVD, similar to that of  B2108+213, would place them at the high mass end of group scale structures, of which the CNOC2 groups reported by \citep{Park05} are a representative sample. The velocity distributions of the observed member galaxies in our MOS spectroscopy do not show any bimodal signature; our findings have to be modulated by the limited number of confirmed member galaxies, especially in the case of SL2SJ143139, as well as by the errors associated with our redshift measurements using medium resolution spectroscopy. With this caveat, our observations indicate that both our groups are fully virialized, unlike the ongoing merger \citet{McKe09} propose for B2108+213. This high degree of virialization in our groups lends further support to our finding that the density distribution in both these structures resemble the dark matter dominated NFW or Hernquist profiles; our likelihood estimator rejects at well over the $3\sigma$ level an isothermal distribution, which \citet{McKe09} find best fits the mass distribution in B2108+213. Given the active merging associated with the assembly of mass in a group environment, the observed density distribution in these structures may be a sensitive function of the degree of virialization. Results from our observations of the remaining lensing galaxy groups in our catalog will therefore not only add to the sample size but also permit us to test this important hypothesis. 


\section{Summary}  \label{DC}

Our overall objective is to understand the assembly of structure at the mass scale of galaxy groups within the $\Lambda$CDM cosmology. Taking an observational approach, our immediate goal is to characterize groups using their virial masses and M/L ratios, and compare these values against corresponding properties observed in clusters and field galaxies. In addition, we also test if the underlying density distribution in groups is indeed NFW-like as predicted by $\Lambda$CDM, or more isothermal as observed in massive field galaxies. 

\indent The approach we have adopted uses joint constraints on the mass derived from strong lensing, and from the dynamics of member galaxies. We obtain these combined constraints by selectively observing galaxy groups with strong lens features discovered in our catalogs of groups and clusters from $161\sqdeg$ of CFHTLS-Wide imaging \citep{Than09}. The lensing mass is estimated from a lens model using the measured geometry of the lensed images in the available CFHTLS imaging, and from our measurement of the spectroscopic redshifts of the source and deflector. Using the observed colors and luminosities of the central galaxies (enclosed within the arc of the lensed image), we account for their baryonic masses using the correlations provided by \citet{Bell01} with the correlation coefficients taken from \citet{Bell03}. The dynamical mass of each group is estimated from the LOSVD of the member galaxies measured through MOS spectroscopy.  

\indent In this paper, we have provided a detailed description of the formalism we have developed to convert these observables into the parameters we use to characterize the mass distribution in galaxy groups, namely, the virial mass, $M_v$, and the concentration index, $c$, of the underlying density distribution. The formalism is built around three popular, parametric density profiles, namely, the NFW and the Hernquist models used for dark matter dominated systems, and the Softened Isothermal Sphere observed in baryon rich galaxies. As part of this work, we have reformulated their common definitions, which involve various parameters specific to each profile, into a consistent set of non-dimensional expressions in terms of $M_v$ and $c$. This reformulation thus permits us to explore the effects on the observed characteristics of each density distribution due to variations in these two parameters; more importantly, these tests assess how well the observations permit us to distinguish between the density distributions. Our results indicate that estimates of the mass in the \emph{core} of the galaxy group, ($r/r_{vir}\leq0.2$) are essential; since strong lensing is sensitive to the mass distribution in this region, this result offers key support to our observational approach of using groups with lensed arcs as ideal probes.  

\indent The results presented here are based on our pilot Gemini GMOS-MOS observations of two lensing groups, drawn from a present sample of nine candidates in our CFHTLS-Wide lens catalog. The observed LOSVD of SL2SJ143000+554648 (deflector redshift, z=0.501, source z=1.435) is 720$\kms$ determined from 20 confirmed members within a rest frame aperture of $0.77 \mpc$. The significant number of confirmed members permits us to estimate a LOSVD = 784$\kms$ within a rest frame aperture of $0.20 \mpc$ using 6 members; this provides an additional constraint for estimating the virial mass. SL2SJ143139+553323 (deflector z=0.669, source photometric z = 2.083) has an observed LOSVD = 563$\kms$ measured from 9 confirmed members within a rest frame aperture of $0.912 \mpc$. 

\indent The lens model of SL2SJ143000  indicates a total gravitating mass within the Einstein radius of $(6.044\pm0.018)\times10^{12}\msun$, of which $(0.452\pm0.011)\times10^{12}\msun$ is the baryonic contribution from the central galaxies. The corresponding values for SL2SJ143139 are $6.537\pm0.163$ and $0.507\pm0.037$, both given in units of $10^{12}\msun$. Using our formalism, we have assessed that an under- or over-estimate up to 50\% in the lensing mass within the Einstein radius has only a marginal influence ($\leq5\%$) on the virial mass determined by our model. 

\indent These observed lensing and dynamical mass estimates for each group are converted into corresponding $M_v$ and $c$ for each density distribution using a maximum likelihood estimator in conjunction with our models. The estimated virial masses, in units of $10^{14}\msun$ are  $2.8\pm0.4$ for SL2SJ143000, and $1.6\pm0.4$ for SL2SJ143139, for both the NFW and the Hernquist profiles. However, the isothermal sphere was rejected at well over the $3\sigma$ level by the fitting procedure for both these groups. 

\indent The total $i$-band luminosity of the member galaxies summed over an aperture equal to the virial radius, $r_v$, corresponding to the virial mass is used to estimate the M/L ratios of these groups. Member galaxies were selected using a color selection of $\pm0.25$mag relative to the color of the central galaxies; only galaxies brighter than $i=22.5$mag were included to avoid large uncertainties due to increased photometric redshifts at fainter magnitudes. The $i$-band M/L ratio of both groups is $\sim265\pm25$, with only small variations, much less than the associated uncertainties, seen between the NFW and the Hernquist profiles.  

\indent The estimated M/L ratios indicate that both these objects are group scale objects though their virial masses and LOSVD put them at the high mass end of structures in this category. Their density distributions are best fit by dark matter dominated NFW or Hernquist profiles similar to the clusters observed by \citet{Rine06}. Their degree of virialization is also indicated by the unimodal, Gaussian distribution of the observed velocities of their member galaxies, unlike the merging group with an isothermal mass profile and bimodal velocity distribution reported by \citet{McKe09}. Results from our ongoing observations of the remaining lensing galaxy groups in our catalog, will not only provide the characteristics of a greater sample of objects in this important transition range, but will also investigate the effect of virialization on the underlying density distribution of galaxy groups.


\acknowledgments
The results reported here are based on observations obtained at the Gemini Observatory, which is operated by the Association of Universities for Research in Astronomy, Inc., under a cooperative agreement with the NSF on behalf of the Gemini partnership: the National Science Foundation (United States), the Science and Technology Facilities Council (United Kingdom), the National Research Council (Canada), CONICYT (Chile), the Australian Research Council (Australia), MinistŽrio da Cincia e Tecnologia (Brazil) and SECYT (Argentina). This work is also based in part on data products produced at TERAPIX as part of the Canada-France-Hawaii Telescope Legacy Survey, a collaborative project of NRC and CNRS. In addition, this research used the facilities of the Canadian Astronomy Data Centre operated by the National Research Council of Canada with the support of the Canadian Space Agency. The work reported here forms part of a thesis dissertation during which KT was supported by a University of Victoria Graduate Fellowship and a National Research Council of Canada Graduate Student Scholarship Supplement Program (NRC-GSSSP) Award \\


{\it Facilities:} \facility{Gemini}, \facility{Terapix}, \facility{CADC}, \facility{CFHT}. \\



\bibliographystyle{apj}
\bibliography{GrpLens}

\begin{thebibliography}{52}
\providecommand{\natexlab}[1]{#1}
\providecommand{\url}[1]{\texttt{#1}}
\expandafter\ifx\csname urlstyle\endcsname\relax
  \providecommand{\doi}[1]{doi: #1}\else
  \providecommand{\doi}{doi: \begingroup \urlstyle{rm}\Url}\fi

\bibitem[{Bardeau} et~al.(2007){Bardeau}, {Soucail}, {Kneib}, {Czoske},
  {Ebeling}, {Hudelot}, {Smail}, \& {Smith}]{Bard07}
S.~{Bardeau}, G.~{Soucail}, J.~{Kneib}, O.~{Czoske}, H.~{Ebeling},
  P.~{Hudelot}, I.~{Smail}, \& G.~P. {Smith}.
\newblock 
\newblock 2007, \emph{\aap}, 470, 449.

\bibitem[{Beers} et~al.(1990){Beers}, {Flynn}, \& {Gebhardt}]{Timo90}
T.~C. {Beers}, K.~{Flynn}, \& K.~{Gebhardt}.
\newblock  
\newblock 1990, \emph{\aj}, 100, 849.

\bibitem[{Beers} et~al.(1991){Beers}, {Gebhardt}, {Forman}, {Huchra}, \&
  {Jones}]{Beer91}
T.~C. {Beers}, K.~{Gebhardt}, W.~{Forman}, J.~P. {Huchra}, \& C.~{Jones}.
\newblock 
\newblock 1991, \emph{\aj}, 102, 1581.

\bibitem[{Bell} \& {de Jong}(2001)]{Bell01}
E.~F. {Bell} \& R.~S. {de Jong}.
\newblock 
\newblock 2001, \emph{\apj}, 550, 212.

\bibitem[{Bell} et~al.(2003){Bell}, {McIntosh}, {Katz}, \& {Weinberg}]{Bell03}
E.~F. {Bell}, D.~H. {McIntosh}, N.~{Katz}, \& M.~D. {Weinberg}.
\newblock 
\newblock 2003, \emph{\apjs}, 149, 289.

\bibitem[{Binney}(1980)]{Binn80}
J.~{Binney}.
\newblock 
\newblock 1980, \emph{\mnras}, 190, 873.

\bibitem[{Binney} \& {Tremaine}(1987)]{Binn87}
J.~{Binney} \& S.~{Tremaine}.
\newblock 
\newblock 1987, \emph{{Galactic dynamics}}, (Princeton, NJ, Princeton University Press).

\bibitem[{Blanton} et~al.(2003){Blanton}, {Hogg}, {Bahcall}, {Brinkmann},
  {Britton}, {Connolly}, {Csabai}, {Fukugita}, {Loveday}, {Meiksin}, {Munn},
  {Nichol}, {Okamura}, {Quinn}, {Schneider}, {Shimasaku}, {Strauss}, {Tegmark},
  {Vogeley}, and {Weinberg}]{Blan03}
M.~R. {Blanton}, et~al.
\newblock 
\newblock 2003, \emph{\apj}, 592, 819.

\bibitem[{Bolzonella} et~al.(2000){Bolzonella}, {Miralles}, and
  {Pell{\'o}}]{Bolz00}
M.~{Bolzonella}, J.-M. {Miralles}, \& R.~{Pell{\'o}}.
\newblock 
\newblock 2000, \emph{\aap}, 363, 476.

\bibitem[{Broadhurst} et~al.(2008){Broadhurst}, {Umetsu}, {Medezinski},
  {Oguri}, \& {Rephaeli}]{Broa08}
T.~{Broadhurst}, K.~{Umetsu}, E.~{Medezinski}, M.~{Oguri}, \& Y.~{Rephaeli}.
\newblock 
\newblock 2008, \emph{\apjl}, 685, 9.

\bibitem[{Cabanac} et~al.(2007){Cabanac}, {Alard}, {Dantel-Fort}, {Fort},
  {Gavazzi}, {Gomez}, {Kneib}, {Le F{\`e}vre}, {Mellier}, {Pello}, {Soucail},
  {Sygnet}, \& {Valls-Gabaud}]{Caba07}
R.~A. {Cabanac}, et~al.
\newblock 
\newblock 2007, \emph{\aap}, 461, 813.

\bibitem[{Cava} et~al.(2009){Cava}, {Bettoni}, {Poggianti}, {Couch}, {Moles},
  {Varela}, {Biviano}, {D'Onofrio}, {Dressler}, {Fasano}, {Fritz},
  {Kj{\ae}rgaard}, {Ramella}, \& {Valentinuzzi}]{Cava09}
A.~{Cava}, et~al.
\newblock 
\newblock 2009, \emph{\aap}, 495, 707.

\bibitem[{Cole} \& {Lacey}(1996)]{Cole96}
S.~{Cole} \& C.~{Lacey}.
\newblock 
\newblock 1996, \emph{\mnras}, 281, 716.

\bibitem[{Comerford} \& {Natarajan}(2007)]{Come07}
J.~M. {Comerford} \& P.~{Natarajan}.
\newblock 
\newblock 2007, \emph{\mnras}, 379, 190.

\bibitem[{Eke} et~al.(2004){Eke}, {Baugh}, {Cole}, {Frenk}, {Norberg},
  {Peacock}, {Baldry}, {Bland-Hawthorn}, {Bridges}, {Cannon}, {Colless},
  {Collins}, {Couch}, {Dalton}, {de Propris}, {Driver}, {Efstathiou}, {Ellis},
  {Glazebrook}, {Jackson}, {Lahav}, {Lewis}, {Lumsden}, {Maddox}, {Madgwick},
  {Peterson}, {Sutherland}, \& {Taylor}]{Eke04a}
V.~R. {Eke}, et~al.
\newblock 
\newblock 2004, \emph{\mnras}, 348, 866.

\bibitem[{Fassnacht} et~al.(2008){Fassnacht}, {Kocevski}, {Auger}, {Lubin},
  {Neureuther}, {Jeltema}, {Mulchaey}, \& {McKean}]{Fass08}
C.~D. {Fassnacht}, D.~D. {Kocevski}, M.~W. {Auger}, L.~M. {Lubin}, J.~L.
  {Neureuther}, T.~E. {Jeltema}, J.~S. {Mulchaey}, \& J.~P. {McKean}.
\newblock 
\newblock 2008, \emph{\apj}, 681, 1017.

\bibitem[{Freeman}(2001)]{Free01}
K.~C. {Freeman}.
\newblock 
\newblock 2001, in ASP Conf. Ser. 240, \emph{Gas and Galaxy Evolution}, ed. J.~E. {Hibbard},  M.~{Rupen}, \& J.~H. {van Gorkom}, (San Francisco, CA:ASP), 240.

\bibitem[{Gavazzi} et~al.(2008){Gavazzi}, {Treu}, {Koopmans}, {Bolton},
  {Moustakas}, {Burles}, \& {Marshall}]{Gava08}
R.~{Gavazzi}, T.~{Treu}, L.~V.~E. {Koopmans}, A.~S. {Bolton}, L.~A.
  {Moustakas}, S.~{Burles}, \& P.~J. {Marshall}.
\newblock 
\newblock 2008, \emph{\apj}, 677, 1046.

\bibitem[{Gebhardt} \& {Beers}(1991)]{Gebh91}
K.~{Gebhardt} \& T.~C. {Beers}.
\newblock 
\newblock 1991, \emph{\apj}, 383, 72.

\bibitem[{Hartley} et~al.(2008){Hartley}, {Gazzola}, {Pearce}, {Kay}, \&
  {Thomas}]{Hart08}
W.~G. {Hartley}, L.~{Gazzola}, F.~R. {Pearce}, S.~T. {Kay}, \& P.~A. {Thomas}.
\newblock 
\newblock 2008, \emph{\mnras}, 386, 2015.

\bibitem[{Helsdon} \& {Ponman}(2000)]{Hels00}
S.~F. {Helsdon} \& T.~J. {Ponman}.
\newblock 
\newblock 2000, \emph{\mnras}, 315, 356.

\bibitem[{Hernquist}(1990)]{Hern90}
L.~{Hernquist}.
\newblock 
\newblock 1990, \emph{\apj}, 356, 359.

\bibitem[{Hoekstra} et~al.(2001){Hoekstra}, {Franx}, {Kuijken}, {Carlberg},
  {Yee}, {Lin}, {Morris}, {Hall}, {Patton}, {Sawicki}, \& {Wirth}]{Hoek01}
H.~{Hoekstra}, et~al.
\newblock 
\newblock 2001, \emph{\apjl}, 548, 5.

\bibitem[{Kelson}(2003)]{Kels03}
D.~D. {Kelson}.
\newblock 
\newblock 2003, \emph{\pasp}, 115, 688.

\bibitem[{Kinney} et~al.(1996){Kinney}, {Calzetti}, {Bohlin}, {McQuade},
  {Storchi-Bergmann}, \& {Schmitt}]{Kinn96}
A.~L. {Kinney}, D.~{Calzetti}, R.~C. {Bohlin}, K.~{McQuade},
  T.~{Storchi-Bergmann}, \& H.~R. {Schmitt}.
\newblock 
\newblock 1996, \emph{\apj}, 467, 38.

\bibitem[{Koester} et~al.(2007){Koester}, {McKay}, {Annis}, {Wechsler},
  {Evrard}, {Bleem}, {Becker}, {Johnston}, {Sheldon}, {Nichol}, {Miller},
  {Scranton}, {Bahcall}, {Barentine}, {Brewington}, {Brinkmann}, {Harvanek},
  {Kleinman}, {Krzesinski}, {Long}, {Nitta}, {Schneider}, {Sneddin}, {Voges},
  and {York}]{Koes07}
B.~P. {Koester}, et~al.
\newblock 
\newblock 2007,\emph{\apj}, 660, 239.

\bibitem[{Koopmans} et~al.(2006){Koopmans}, {Treu}, {Bolton}, {Burles}, \&
  {Moustakas}]{Koop06}
L.~V.~E. {Koopmans}, T.~{Treu}, A.~S. {Bolton}, S.~{Burles}, \& L.~A.
  {Moustakas}.
\newblock 
\newblock 2006, \emph{\apj}, 649, 599.

\bibitem[{Kravtsov} et~al.(1998){Kravtsov}, {Klypin}, {Bullock}, \&
  {Primack}]{Krav98}
A.~V. {Kravtsov}, A.~A. {Klypin}, J.~S. {Bullock}, \& J.~R. {Primack}.
\newblock 
\newblock 1998, \emph{\apj}, 502, 48.

\bibitem[{Kubo} et~al.(2009){Kubo}, {Allam}, {Annis}, {Buckley-Geer}, {Diehl},
  {Kubik}, {Lin}, \& {Tucker}]{Kubo09}
J.~M. {Kubo}, S.~S. {Allam}, J.~{Annis}, E.~J. {Buckley-Geer}, H.~T. {Diehl},
  D.~{Kubik}, H.~{Lin}, \& D.~{Tucker}.
\newblock 
\newblock 2009, \emph{\apjl}, 696, 61.

\bibitem[{Limousin} et~al.(2009){Limousin}, {Cabanac}, {Gavazzi}, {Kneib},
  {Motta}, {Richard}, {Thanjavur}, {Foex}, {Pello}, {Crampton}, {Faure},
  {Fort}, {Jullo}, {Marshall}, {Mellier}, {More}, {Soucail}, {Suyu},
  {Swinbank}, {Sygnet}, {Tu}, {Valls-Gabaud}, {Verdugo}, \& {Willis}]{Limo09}
M.~{Limousin}, et~al.
\newblock 
\newblock 2009, \emph{\aap}, 502, 445.

\bibitem[{{\L}okas} \& {Mamon}(2001)]{Loka01}
E.~L. {{\L}okas} \& G.~A. {Mamon}.
\newblock 
\newblock 2001, \emph{\mnras}, 321, 155.

\bibitem[{Lopes} et~al.(2009){Lopes}, {de Carvalho}, {Kohl-Moreira}, \&
  {Jones}]{Lope09}
P.~A.~A. {Lopes}, R.~R. {de Carvalho}, J.~L. {Kohl-Moreira}, \& C.~{Jones}.
\newblock 
\newblock 2009, \emph{\mnras}, 392, 135.

\bibitem[{Marinoni} \& {Hudson}(2002)]{Mari02}
C.~{Marinoni} \& M.~J. {Hudson}.
\newblock 
\newblock 2002, \emph{\apj}, 569, 101.

\bibitem[{Markevitch}(1998)]{Mark98}
M.~{Markevitch}.
\newblock 
\newblock 1998, \emph{\apj}, 504, 27.

\bibitem[{McKean} et~al.(2009){McKean}, {Auger}, {Koopmans}, {Vegetti},
  {Czoske}, {Fassnacht}, {Treu}, {More}, \& {Kocevski}]{McKe09}
J.~P. {McKean}, et~al.
\newblock 
\newblock 2009, \emph{ArXiv 0910.1133}.

\bibitem[{McQuade} et~al.(1995){McQuade}, {Calzetti}, \& {Kinney}]{Mcqu95}
K.~{McQuade}, D.~{Calzetti}, \& A.~L. {Kinney}.
\newblock 
\newblock 1995, \emph{\apjs}, 97, 331.

\bibitem[{Meneghetti} et~al.(2007){Meneghetti}, {Bartelmann}, {Jenkins}, \&
  {Frenk}]{Mene07}
M.~{Meneghetti}, M.~{Bartelmann}, A.~{Jenkins}, \& C.~{Frenk}.
\newblock 
\newblock 2007, \emph{\mnras}, 381, 171.

\bibitem[{Mulchaey}(2004)]{Mulc04}
J.~S. {Mulchaey}.
\newblock 
\newblock 2004, in \emph{Clusters of Galaxies: Probes of Cosmological Structure and Galaxy
  Evolution}, ed. J.~S. {Mulchaey}, A.~{Dressler}, \& A.~{Oemler}, (Cambridge Univ. Press).

\bibitem[{Navarro} et~al.(1997){Navarro}, {Frenk}, \& {White}]{Nava97}
J.~F. {Navarro}, C.~S. {Frenk}, \& S.~D.~M. {White}.
\newblock 
\newblock 1997, \emph{\apj}, 490, 493.

\bibitem[{Osmond} \& {Ponman}(2004)]{Osmo04}
J.~P.~F. {Osmond} \& T.~J. {Ponman}.
\newblock 
\newblock 2004, \emph{\mnras}, 350, 1511.

\bibitem[{Parker} et~al.(2005){Parker}, {Hudson}, {Carlberg}, \&
  {Hoekstra}]{Park05}
L.~C. {Parker}, M.~J. {Hudson}, R.~G. {Carlberg}, \& H.~{Hoekstra}.
\newblock 
\newblock 2005, \emph{\apj}, 634, 806.

\bibitem[{Piessens} et~al.(1983){Piessens}, {de Doncker-Kapenga}, \&
  {Ueberhuber}]{Pies83}
R.~{Piessens}, E.~{de Doncker-Kapenga}, \& C.~W. {Ueberhuber}.
\newblock 
\newblock 1983, \emph{{Quadpack. A subroutine package for automatic integration}}, (Springer Series in Computational Mathematics, Berlin: Springer).

\bibitem[{Press} et~al.(1992){Press}, {Teukolsky}, {Vetterling}, \&
  {Flannery}]{Pres92}
W.~H. {Press}, S.~A. {Teukolsky}, W.~T. {Vetterling}, \& B.~P. {Flannery}.
\newblock 
\newblock 1992, \emph{{Numerical recipes in C. The art of scientific computing}}, (Cambridge: University Press).

\bibitem[{Rasmussen} et~al.(2006){Rasmussen}, {Ponman}, {Mulchaey}, {Miles},
  \& {Raychaudhury}]{Rasm06}
J.~{Rasmussen}, T.~J. {Ponman}, J.~S. {Mulchaey}, T.~A. {Miles}, \&
  S.~{Raychaudhury}.
\newblock 
\newblock 2006,\emph{\mnras}, 373, 653.

\bibitem[{Rines} \& {Diaferio}(2006)]{Rine06}
K.~{Rines} \& A.~{Diaferio}.
\newblock 
\newblock 2006, \emph{\aj}, 132, 1275.

\bibitem[{Schneider} et~al.(1992){Schneider}, {Ehlers}, and {Falco}]{Schn92}
P.~{Schneider}, J.~{Ehlers}, and E.~E. {Falco}.
\newblock \emph{{Gravitational Lenses}}.
\newblock Springer-Verlag New York.

\bibitem[{Sheldon} et~al.(2009){Sheldon}, {Johnston}, {Masjedi}, {McKay},
  {Blanton}, {Scranton}, {Wechsler}, {Koester}, {Hansen}, {Frieman}, and
  {Annis}]{Shel09b}
E.~S. {Sheldon}, D.~E. {Johnston}, M.~{Masjedi}, T.~A. {McKay}, M.~R.
  {Blanton}, R.~{Scranton}, R.~H. {Wechsler}, B.~P. {Koester}, S.~M. {Hansen},
  J.~A. {Frieman}, and J.~{Annis}.
\newblock 
\newblock 2009, \emph{\apj}, 703, 2232.

\bibitem[{Smith} et~al.(2005){Smith}, {Kneib}, {Smail}, {Mazzotta}, {Ebeling},
  \& {Czoske}]{Smit05a}
G.~P. {Smith}, J.-P. {Kneib}, I.~{Smail}, P.~{Mazzotta}, H.~{Ebeling}, \&
  O.~{Czoske}.
\newblock 
\newblock 2005, \emph{\mnras}, 359, 417

\bibitem[{Tago} et~al.(2006){Tago}, {Einasto}, {Saar}, {Einasto}, {Suhhonenko},
  {Joeveer}, {Vennik}, {Heinamaki}, \& {Tucker}]{Tago06}
E.~{Tago}, et~al.
\newblock 2006, \emph{AN}, 327, 365.

\bibitem[{Tago} et~al.(2008){Tago}, {Einasto}, {Saar}, {Tempel}, {Einasto},
  {Vennik}, \& {M{\"u}ller}]{Tago08}
E.~{Tago}, J.~{Einasto}, E.~{Saar}, E.~{Tempel}, M.~{Einasto}, J.~{Vennik}, \&
  V.~{M{\"u}ller}.
\newblock 
\newblock 2008, \emph{\aap}, 479, 927.

\bibitem[{Thanjavur} et~al.(2009){Thanjavur}, {Willis}, \&
  {Crampton}]{Than09}
K.~{Thanjavur}, J.~{Willis}, \& D.~{Crampton}.
\newblock 
\newblock 2009, \emph{\apj}, 706, 571.

\bibitem[{Tonry} \& {Davis}(1979)]{Tonr79}
J.~{Tonry} \& M.~{Davis}.
\newblock 
\newblock 1979, \emph{\aj}, 84, 1511.

\bibitem[{Treu} \& {Koopmans}(2004)]{Treu04}
T.~{Treu} \& L.~V.~E. {Koopmans}.
\newblock 
\newblock 2004, \emph{\apj}, 611, 739.

\bibitem[{Tu} et~al.(2008){Tu}, {Limousin}, {Fort}, {Shu}, {Sygnet}, {Jullo},
  {Kneib}, \& {Richard}]{Tu08}
H.~{Tu}, M.~{Limousin}, B.~{Fort}, C.~G. {Shu}, J.~F. {Sygnet}, E.~{Jullo},
  J.~P. {Kneib}, \& J.~{Richard}.
\newblock 
\newblock 2008, \emph{\mnras}, 386, 1169.

\bibitem[{Willick} \& {Padmanabhan}(2000)]{Will00a}
J.~A. {Willick} \& N.~{Padmanabhan}.
\newblock 
\newblock 2000, \emph{ArXiv:astro-ph/0012253}.

\bibitem[{Willis} et~al.(2005){Willis}, {Pacaud}, {Valtchanov}, {Pierre},
  {Ponman}, {Read}, {Andreon}, {Altieri}, {Quintana}, {Dos Santos},
  {Birkinshaw}, {Bremer}, {Duc}, {Galaz}, {Gosset}, {Jones}, \&
  {Surdej}]{Will05}
J.~P. {Willis}, et~al.
\newblock 
\newblock 2005, \emph{\mnras}, 363, 675.

\end{thebibliography}
 
\clearpage

\clearpage


\clearpage


\clearpage


\clearpage


\clearpage


\clearpage


\clearpage


\clearpage


\clearpage


\begin{table*}[htpb]
\caption{Modeled values of projected mass and LOSVD as functions of virial mass}
\begin{center}
\scriptsize
\begin{tabular}{|c|c|c|c|c|c|c|c|c|c|}
\hline
$M_v$ &  $r_v$ &  $V_v$ &  $s_E$ & \multicolumn{3}{|c|}{$M_p \; [10^{12}\msun$]} &  \multicolumn{3}{|c|}{$\sigma_{LOS} \; [\kms]$}  \\
\hline
[$10^{14}\msun$] &  [$\mpc$] &  [$\kms$] &  [-] & NFW & HRQ & SIS & NFW & HRQ & SIS  \\

\hline
10.  &      2.09     &     1434.   &       0.014   &      5.67   &      5.84    &     5.65      &     907.      &     889.     &     1011.  \\
7.5    &     1.90     &     1303.   &       0.015   &      4.76    &     4.90    &     4.71    &      815.      &     794.     &      921.  \\
5.0    &     1.66     &     1138.   &       0.017   &      3.64   &      3.74   &      3.63    &       696.      &     673.     &      806.  \\
2.5    &     1.32      &     904.   &       0.022    &     2.21   &      2.23    &     2.31     &      526.     &      498.     &      640.  \\
1.0   &      0.97      &     666.   &       0.030    &     1.04    &     1.01    &     1.25     &      353.      &     324.      &     472.  \\
\hline
\end{tabular}
\end{center}
{\footnotesize Modeled values of projected mass and aperture LOSVD as functions of virial mass, $M_v$, for the NFW, HRQ and SIS density distributions (column 1). The corresponding halo properties tabulated are virial radius, $r_v$, circular velocity, $V_v$, scaled Einstein radius, $s_E$, the projected mass, $M_p$ and observed LOSVD, $\sigma_{LOS}$, for NFW, HRQ and SIS respectively; the listed projected mass and LOSVD are interpolated values corresponding to the Einstein radius (=29Kpc) and observed aperture radius (=0.77 Mpc) of the galaxy group SL2SJ143000 (see Section \ref{LOSres2} for details)}
\label{Tbl1}
\end{table*}
\normalsize 

\clearpage


\begin{table*}[ph]
\caption{Details of confirmed cluster members in SL2SJ143000+554648}
\begin{center}
\scriptsize
\begin{tabular}{|c|c|c|c|c|c|c|c|c|c|}
\hline
ID &  RA &  Dec &  i &  (g-r) & (r-i)  & z  &  $\Delta z$ & RA offset & Dec offset \\
- & deg. & deg. & mag & mag & mag & - &  $\kms$ & arc sec & arc sec \\
\hline
\multicolumn{10}{|c|}{\textbf{SL2SJ143000} }  \\
\hline
BCG   &      217.502833   &       55.779878   &     19.049   &      1.638   &      0.880   &  0.4972 &   112.4   &     0.0   &     0.0   \\
 \#1   &      217.559250   &       55.764219   &      19.43   &       1.43   &       0.70   &  0.5038 &    72.8   &   203.1   &   -56.4   \\
 \#2   &      217.456500   &       55.769869   &      20.09   &       1.55   &       0.81   &  0.4959 &   131.0   &  -166.8   &   -36.0   \\
 \#3   &      217.446750   &       55.768939   &      20.10   &       1.65   &       0.92   &  0.4992 &   140.3   &  -201.9   &   -39.4   \\
 \#4   &      217.550250   &       55.756489   &      20.32   &       1.71   &       0.87   &  0.5052 &    99.2   &   170.7   &   -84.2   \\
 \#5   &      217.445417   &       55.767169   &      20.36   &       1.59   &       0.79   &  0.4988 &   128.6   &  -206.7   &   -45.8   \\
 \#6   &      217.504583   &       55.780339   &      20.48   &       1.64   &       0.89   &  0.5009 &   170.3   &     6.3   &     1.7   \\
 \#7   &      217.555208   &       55.792289   &      20.69   &       1.58   &       0.84   &  0.5024 &   109.4   &   188.6   &    44.7   \\
 \#8   &      217.497292   &       55.775089   &      20.77   &       1.74   &       0.90   &  0.5019 &    82.4   &   -20.0   &   -17.2   \\
 \#9   &      217.506000   &       55.788300   &      20.80   &       1.61   &       0.84   &  0.4969 &   126.8   &    11.4   &    30.3   \\
\#10   &      217.561667   &       55.765411   &      20.80   &       1.42   &       0.65   &  0.5081 &   117.8   &   211.8   &   -52.1   \\
\#11   &      217.515917   &       55.782339   &      20.80   &       1.69   &       0.87   &  0.4952 &   120.5   &    47.1   &     8.9   \\
\#12   &      217.459958   &       55.761981   &      20.81   &       1.44   &       0.69   &  0.4987 &   120.5   &  -154.4   &   -64.4   \\
\#13   &      217.513917   &       55.772692   &      21.04   &       1.66   &       0.90   &  0.5018 &   128.3   &    39.9   &   -25.9   \\
\#14   &      217.516000   &       55.773239   &      21.17   &       1.42   &       0.67   &  0.5018 &   115.1   &    47.4   &   -23.9   \\
\#15   &      217.507792   &       55.780000   &      21.20   &       1.42   &       0.70   &  0.5041 &    79.4   &    17.8   &     0.4   \\
\#16   &      217.557292   &       55.794850   &      21.23   &       1.54   &       0.68   &  0.5017 &   140.0   &   196.0   &    53.9   \\
\#17   &      217.533917   &       55.763511   &      21.25   &       1.47   &       0.72   &  0.4990 &   174.5   &   111.9   &   -58.9   \\
\#18   &      217.545917   &       55.756819   &      21.31   &       1.44   &       0.69   &  0.5047 &   114.5   &   155.1   &   -83.0   \\
\#19   &      217.497167   &       55.759511   &      21.34   &       1.41   &       0.65   &  0.5017 &   142.1   &   -20.4   &   -73.3   \\
\hline
\hline
\multicolumn{10}{|c|}{\textbf{SL2SJ143139} }  \\
\hline
BCG   &      217.915500   &       55.556389   &     18.917   &      1.512   &      0.960   &  0.6671 &    84.2   &     0.0   &     0.0   \\
 \#1   &      217.878583   &       55.551211   &      20.48   &       1.48   &       1.01   &  0.6690 &   203.3   &  -132.9   &   -18.6   \\
 \#2   &      217.978333   &       55.565781   &      20.55   &       1.28   &       0.77   &  0.6722 &    95.6   &   226.2   &    33.8   \\
 \#3   &      217.920000   &       55.570400   &      20.70   &       1.55   &       0.98   &  0.6747 &   150.5   &    16.2   &    50.4   \\
 \#4   &      217.872750   &       55.562561   &      21.06   &       1.49   &       1.14   &  0.6694 &    70.2   &  -153.9   &    22.2   \\
 \#5   &      217.886250   &       55.553411   &      21.07   &       1.43   &       0.93   &  0.6667 &   152.6   &  -105.3   &   -10.7   \\
 \#6   &      217.933333   &       55.557511   &      21.07   &       1.37   &       1.15   &  0.6667 &   164.0   &    64.2   &     4.0   \\
 \#7   &      217.950917   &       55.553389   &      21.14   &       1.41   &       0.88   &  0.6702 &    92.3   &   127.5   &   -10.8   \\
 \#8   &      217.912333   &       55.570531   &      21.32   &       1.71   &       1.10   &  0.6731 &   149.9   &   -11.4   &    50.9   \\
\hline
\end{tabular}
\end{center}
{\footnotesize Details of the spectroscopically confirmed cluster members of lensing group SL2SJ143000 and SL2SJ143139. The columns tabulated are member ID, sky position (J2000 $\alpha, \delta$), i magnitude, (g-r) and (r-i) colors, spectroscopic redshift, redshift error, positional offsets in RA and Dec measured relative to the position of the BCG; the numbers annotated on the corresponding panels in Figure \ref{Fig7} follow the order in this table, beginning with 1 for the BCG.}
\label{Tbl2}
\end{table*}
\normalsize 

\clearpage


\begin{table*}[ph]
\caption{\emph{ \emph{ROSTAT}} results of the LOSVD for two lensing groups}
\begin{center}
\scriptsize
\begin{tabular}{|c|c|c|c|c|c|c|}
\hline
SL2S & Bi-weight & BCG &  \multicolumn{2}{|c|}{Rest frame LOSVD [$\kms$]} &  \multicolumn{2}{|c|}{Aperture radius [$\mpc$]}  \\
ID & mean z & z & Inner & Outer & Inner & Outer  \\
\hline
SL2SJ143000  &  $0.5008^{+0.0013}_{-0.001}$  &  $0.497\pm0.001$  & $784^{+126}_{-115}$  & $720^{+91}_{-110}$ & 0.2 & 0.77  \bigstrut  \\
\hline
SL2SJ143139  &  $0.6697^{+0.0023}_{-0.002}$  & $0.667\pm0.001$ & - & $563^{+87}_{-137}$ & - & 0.912  \bigstrut  \\
\hline
\end{tabular}
\end{center}
{\footnotesize Median redshift and LOSVD of confirmed members of the two lensing groups observed by our Gemini GMOS-MOS campaign; the statistical values tabulated are the bi-weight central location and spread computed by \emph{ROSTAT} using the spectroscopic redshifts and errors. The reported $1\sigma$ confidence intervals on the group redshift and the rest frame LOSVD are bootstrap values from  \emph{ROSTAT}, converted to rest frame values using the redshift of the group}
\label{Tbl3}
\end{table*}
\normalsize 

\clearpage

\begin{table*}[ph]
\caption{Lensing geometry and corresponding lensing mass estimates}
\begin{center}
\scriptsize
\begin{tabular}{|c|c|c|c|c|c|c|}
\hline
SL2S ID & $z_d$ & $z_s$ & $\theta_E$ & $\mathrm{LM_{tot}}$ & $\mathrm{LM_{bary}}$ & $\mathrm{LM_{DM}}$ \\ 
\hline
SL2SJ143000  & $0.501\pm0.001$ &  $1.435\pm0.001$  & $4\asec.6$  & $6.044\pm0.018$  & $0.427\pm0.011$ & $5.616\pm0.021$   \bigstrut  \\
\hline
SL2SJ143139  & $0.669\pm0.001$ & $\mathit{2.083\pm0.07}$ & $4\asec.35$   & $6.537\pm0.163$ & $0.507\pm0.037$  & $6.03\pm0.167$   \bigstrut  \\
\hline
\end{tabular}
\end{center}
{\footnotesize Estimated lensing masses, $\mathrm{LM_{tot}}$ [$10^{12}\msun$] and associated uncertainties in the two groups from our lens model, using the measured Einstein radius, $\theta_E\;[\asec]$, and the redshifts of the deflector,  $z_d$ and source, $z_s$. The baryonic contribution, $\mathrm{LM_{bary}}$ from Eqn \ref{clr2ML} and the net dark matter lensing mass, $\mathrm{LM_{DM}}$ and corresponding uncertainties, in units of $10^{12}\msun$ are listed in the last two columns.\\
For the lensing group, SL2SJ143139+553323, the photometric redshift and uncertainty are shown italicized.}
\label{Tbl4}
\end{table*}
\normalsize 

\clearpage

\begin{table*}[htpb]
\caption{Maximum likelihood estimates of the virial mass and concentration index}
\begin{center}
\scriptsize
\begin{tabular}{|c|c|c|c||c|c|c|}
\hline
SL2S ID & \multicolumn{3}{|c||}{Virial mass, $M_v$ [$10^{14}\msun$] } & \multicolumn{3}{|c|}{Concentration Index, $c$ [ - ]}  \\
 & NFW & HRQ & SIS & NFW & HRQ & SIS \\
\hline
SLS2J143000 & $2.78\pm0.38$ & $2.83\pm0.39$ & $3.41\pm0.47$ & $7.24\pm1.32$ & $3.34\pm0.3$ & $0.22\pm0.12$   \bigstrut  \\
\hline
SLS2J143139 & $1.56\pm0.35$ & $1.73\pm0.37$ & $1.17\pm0.58$ & $9.02\pm1.89$ & $3.79\pm0.34$ & $0.34\pm0.18$   \bigstrut  \\
\hline
\end{tabular}
\end{center}
{\footnotesize Maximum likelihood estimates and corresponding uncertainties of the virial mass of each observed galaxy group and the concentration index for each of the assumed underlying density distributions}
\label{Tbl5}
\end{table*}
\normalsize 

\clearpage

\begin{table*}[htpb]
\caption{Observed values and modeled estimates of the lensing mass and LOSVD}
\begin{center}
\scriptsize
\begin{tabular}{|c|c|c|c|c||c|c|c|c|}
\hline
SL2S ID & $M_p$ & \multicolumn{3}{|c||}{Estimated $M_p$ } & $LOSVD$ & \multicolumn{3}{|c|}{Estimated LOSVD}  \\
 & Obs. & NFW & HRQ & SIS & Obs. & NFW & HRQ & SIS \\
\hline
SLS2J143000 & 5.616 & 5.615 & 5.609 & 0.627 & 784. & 788.51 & 803.34 & 831.77  \bigstrut[t]  \\
 &  & & & & 720. & 715.81 & 702.14 & 875.29  \bigstrut [b] \\
\hline
SL2SJ143139 & 6.031 & 6.031 & 6.031 & 2.22 & 563. & 563.1 & 564.3 & 694.3  \bigstrut  \\
\hline
\end{tabular}
\end{center}
{\footnotesize Comparison of the observed values of the lensing masses, $M_p$ and the LOSVD of the two groups, and the estimates from our model corresponding to the best fit parameter values for the three density profiles, given in Table \ref{Tbl5}. The projected masses are in units of  $10^{12}\,\msun$, and the LOSVD are given in $\kms$. }
\label{Tbl6}
\end{table*}
\normalsize 

\clearpage

\begin{table*}[htpb]
\caption{Estimated M/L ratio of the two lensing groups}
\begin{center}
\scriptsize
\begin{tabular}{|c|c|c|c|c|c|}
\hline
SL2S ID & N & $L_{mem}$ & $M_{bary,mem}$ & $(M/L)_{NFW}$ & $(M/L)_{HRQ}$ \bigstrut [t] \\
 &  & [$10^{12}\Lsun$] & [$10^{13}\msun$] & $\msun/\Lsun$ & $\msun/\Lsun$ \bigstrut[b] \\
\hline
SL2SJ143000 & 37 & $1.103\pm0.009$ & $1.032\pm0.007$ & $261\pm18$ & $265\pm19$  \bigstrut  \\
\hline
SL2SJ143139 & 15 & $0.645\pm0.007$ & $0.979\pm0.009$ & $258\pm29$ & $285\pm31$  \bigstrut \\
\hline
\end{tabular}
\end{center}
{\footnotesize For each lensing group, the columns listed are the number of member galaxies, N, the total luminosity, L, and baryonic mass, M, of the members and the M/L ratio of the group for the NFW and the HRQ density distribution along with corresponding uncertainties}
\label{Tbl7}
\end{table*}
\normalsize 


\end{document}